\documentclass[conference]{IEEEtran}
\IEEEoverridecommandlockouts
\usepackage{subfigure}
\usepackage{color}
\usepackage{float}
\usepackage{stfloats}
\usepackage{multirow}
\usepackage{amsmath,amssymb,amsfonts}
\usepackage[ruled]{algorithm2e}
\usepackage{graphicx}
\usepackage{textcomp}
\usepackage{xcolor}
\usepackage[style=numeric, sorting=none]{biblatex}
\def\BibTeX{{\rm B\kern-.05em{\sc i\kern-.025em b}\kern-.08em
    T\kern-.1667em\lower.7ex\hbox{E}\kern-.125emX}}

\begin{document}
\title{R-PMAC: A Robust Preamble Based MAC Mechanism Applied in Industrial Internet of Things\\
}

\author{
 \IEEEauthorblockN{Kai Song, Biqian Feng, Yongpeng Wu, Zhen Gao and Wenjun Zhang}

 \thanks{
 K. Song, B. Feng, Y. Wu, W. Zhang are with the Department of Electronic Engineering, Shanghai Jiao Tong University, Minhang 200240, China (e-mail: gansusongkai@sjtu.edu.cn; fengbiqian@sjtu.edu.cn; yongpeng.wu@sjtu.edu.cn; zhangwenjun@sjtu.edu.cn) (Corresponding author: Yongpeng Wu).}
 
 \thanks{
 Z. Gao is with the School of Information and Electronics and the Advanced Research Institute of Multidisciplinary Science, Beijing Institute of Technology, Beijing 100081, China, and also with the Yangtze Delta Region Academy of Beijing Institute of Technology, Jiaxing 314001, China (email: gaozhen16@bit.edu.cn).}

 
 
}

\maketitle

\begin{abstract}
This paper proposes a novel media access control (MAC) mechanism, called the robust preamble-based MAC mechanism (R-PMAC), which can be applied to power line communication (PLC) networks in the context of the Industrial Internet of Things (IIoT). Compared with other MAC mechanisms such as P-MAC and the MAC layer of IEEE1901.1, R-PMAC has higher networking speed. Besides, it supports whitelist authentication and functions properly in the presence of data frame loss. Firstly, we outline three basic mechanisms of R-PMAC, containing precise time difference calculation, preambles generation and short ID allocation. Secondly, we elaborate its networking process of single layer and multiple layers. Thirdly, we illustrate its robust mechanisms, including collision handling and data retransmission. Moreover, a low-cost hardware platform is established to measure the time of connecting hundreds of PLC nodes for the R-PMAC, P-MAC, and IEEE1901.1 mechanisms in a real power line environment.
The experiment results show that R-PMAC outperforms the other mechanisms by achieving a 50\% reduction in networking time. These findings indicate that the R-PMAC mechanism holds great potential for quickly and effectively building a PLC network in actual industrial scenarios.
\end{abstract}


\section{Introduction}
The exponentially growing data produced in manufacturing process and daily life have significantly contributed to human progress. The Internet of Things (IoT), an emerging communication paradigm, offers a promising solution for collecting and managing vast amounts of data, enabling timely detection of changes and informed decision-making. In IoT, the objects of everyday life are equipped with micro-controllers and transceivers for digital communication. With the assistance of suitable protocol stack, they can communicate with other devices or cloud servers \cite{IIoT}. As a predominant branch of IoT, the Industrial Internet of Things (IIoT) focuses on industrial data like temperature, humidity, facility situation and electrical power consumption. Nowadays, the IIoT serves as the foundation of Industry 4.0 and Intelligent Manufacturing \cite{IIoT4_0}, as it updates industrial devices with information technology (IT) systems and eliminates information silos \cite{Silos,Silos2,Silos3}.

In recent years, many extensive research works concentrate
on the increasingly important communication technology for IIoT, including wireless communication technologies such as narrow-band Internet of Things (NB-IoT) and LoRa, as well as wired communication technologies like optical fiber, RS-485, power line communication (PLC).
However, there are many limitations in realizing IIoT, such as the potential blind spots in wireless communication system and the high cost of wired communication system.
In contrast, by utilizing ubiquitous power lines as transmission medium, PLC offers lower deployment costs and stronger signal penetration through walls. Thus, PLC is a promising technology to realize IIoT. In essence, the IIoT is composed of a large number of communication nodes which regularly upload industrial data via low-cost hardware. Hence, in order to implement PLC-based IIoT, we need to design an appropriate protocol stack to connect these nodes with the software applications located on the upper layer of IIoT.

Designing protocols for the physical (PHY) and multi-access control (MAC) layers in PLC-based IIoT network poses some challenges, including the requirement of robustness to varying channel quality, the usage of low-cost hardware devices, and minimizing networking time in large-scale networks. To date, however, existing PHY and MAC layer protocols have not been fully optimized for the specific requirements of PLC-based IIoT. For instance, the home-oriented PLC protocols usually experience performance degradation in industrial scenarios due to significantly reduced available bandwidth. Besides, the existing MAC layer protocols require a significant amount of time to manage the networking and data uploading processes for hundreds of nodes. 
To address these issues, we propose a new MAC layer mechanism called robust preamble-based MAC (R-PMAC) and validate its performance on a practical hardware platform. The main contributions of this paper are summarized as follows:

\begin{itemize}
    \item We propose a novel preamble-based MAC layer protocol called R-PMAC that offers improved robustness and higher networking speeds for PLC-based IIoT. We elaborate on every step of the networking process, in single-layer and multi-layer scenarios, and show that it outperforms both P-MAC and the MAC layer of IEEE1901.1.
    \item To deal with the possible data loss caused by data frame collision and variable channel quality, we develop a collision handling mechanism and a retransmission mechanism. By utilizing these two mechanisms, R-PMAC is capable of completing networking in the presence of data frame loss.
    \item We specify the data frame length, which can simplify the realization of P-MAC. Besides, we meticulously design the structures of various data frames used in R-PMAC to take full advantage of the space of data frame.
    \item We design a practical programmable hardware platform with low-cost chips to compare the proposed R-PMAC with the MAC layer of IEEE1901.1 and P-MAC. 
    The results of experiment demonstrate that the R-PMAC performs better than the other two mechanisms.
\end{itemize}

The rest of this paper is organized as follows. We review the related works in Section II. Section III introduces some basic mechanisms that are necessary for the realization of R-PMAC, while Section IV describes the networking process. In Section V, we propose the mechanisms that can increase the robustness of R-PMAC, followed by the design of the structure of different data frames of R-PMAC in Section VI. We then describe the programmable hardware platform in Section VII, where we implement R-PMAC, P-MAC, and the MAC layer of IEEE1901.1. In Section VIII, we demonstrate the experimental results validating the performance of R-PMAC, and in Section IX, we conclude the paper, highlighting the future development of PLC technology applied in IIoT.

\section{Related work}
The IIoT can be described with a hierarchical model consisting of multiple abstract layers. 
At the upper layers, software applications and interfaces facilitate data exchange, which is similar to traditional computer networks. Common protocols such as HTTP and TCP-IP can be readily adapted to these layers.
Researchers have also developed various IIoT software platforms which runs on servers and focuses on analyzing industrial data \cite{Software}. The lower layers of the IIoT, including physical (PHY) layer and multi-access control (MAC) layer, are responsible for establishing connections between IIoT nodes over unreliable channels. Usually, they need customized designs because their performance determines the data collection rate and the efficiency of network organization and maintenance.

\subsection{The PHY layer of PLC network}
The PHY layer of PLC performs modulation, demodulation, and complicated data processing to transmit data through the PLC channel. In addition to traditional modem techniques like continuous phase frequency shift keying (FSK) \cite{FSK}, researchers attempt to equip PHY layer of PLC with more advanced modulation schemes to further overcome the fading, noise, and multipath propagation present in power lines \cite{multipath}. Orthogonal frequency division multiplexing (OFDM) is such a modulation scheme that can meet the requirement. The OFDM can modulate different carriers individually with a specific multi-bit modulation scheme and signal power allocation using the water-filling principle \cite{BOOK}. In OFDM, the result of the Inverse Fast Fourier Transform (IFFT) of an information vector of length N is transmitted. Multiplying the result at the receiver with the FFT gives back the information vector \cite{BOOK}. OFDM have been pervasively used in both broad-band and narrow-band PLC. For instance, IEEE1901.1 is a classical PLC protocol and is applied in smart home. Baseband data is generated by turbo encoding, scrambling, bit interleaving, and robust OFDM (ROBO) interleaving in the PHY layer of IEEE1901.1, and then modulated via OFDM \cite{IEEE1901}. It can be observed that OFDM is robust to non-AWGN and distance and frequency dependent noise \cite{OFDM}, which indicates that OFDM is suitable for the PHY layer of PLC.

In addition, researchers explore the transplantation of wireless communication techniques to the PHY layer of PLC. For example, multiple lines in a three-phase wire are used to transmit signals, which is similar to the multiple-input-multiple-output (MIMO) structure in wireless communication \cite{MIMO_OFDM}. Besides, wireless and PLC interfaces are integrated in network stations whose control nodes evaluate the channel quality of both wireless and PLC to select the channel with better quality for exchanging data \cite{HYBRID}.

\subsection{The MAC layer of PLC network}
The MAC layer rules the topology of PLC network and the way of nodes joining in the network. Existing protocols like IEEE1901.1 and HomePlug tend to use tree topology or mesh topology according to detailed scenarios \cite{IEEE1901, HOMEPLUG}. In terms of the types of nodes, most protocols use the same classification method, which divides nodes into three categories: Central Coordinators (CCOs), Proxy Coordinators (PCOs), and Stations (STAs). CCO is the unique network gateway, while STAs can serve as PCO on demand when it forwards data between CCO and other STAs.

Two primary objectives of optimizing MAC layer in IIoT are to speed up networking process and handle data frame collisions caused by large number of nodes. MAC protocols can be divided into two main groups: fixed access and dynamic access \cite{BOOK2}. Time Division Multiple Access (TDMA) is a representative fixed access protocol, while dynamic access protocols can be further classified into arbitration protocols (e.g. Token-Passing and Polling protocol) and contention protocols like Carrier Sense Multiple Access (CSMA). In the network based on token-passing and polling protocol, stations exchange token-message, and the station that possesses the token gets the medium \cite{BOOK}. Besides, the active polling scheme, evolved from polling protocol, can keep active network stations and temporarily exclude other stations from the polling cycle, which alleviates the disadvantage of long round-trip time \cite{TOKEN}. Among various contention protocols, CSMA is widely applied in MAC layer to decrease collision. Recent research on CSMA includes modifications of traditional CSMA mechanisms \cite{cluster}, modelling of network topology \cite{Markov}, and compressing the time of channel contention with advanced channel sense technology \cite{MPD}.

As an alternative to CSMA, the novel preamble-based MAC (P-MAC) mechanism is proposed to divide networking process into three stages, Preamble time exchange (PTE), Time query (T-Query), and Network configuration (Net-Config) \cite{PMAC}. PTE involves a downlink preamble broadcast by the CCO and uplink preambles sent by the STAs in response to the CCO. Both CCO and STAs can capture the timing differences between the downlink and uplink preamble. In T-Query and Net-Config, CCO distinguishes STAs with time differences recorded in PTE and control the STAs to join in the network. In order to save time spent on channel contention, P-MAC limits channel contention to the PTE stage and avoid time-consuming collision detection and handling. Since the preambles contain no data, the time slots for preambles in PTE are substantially shorter than those for data frames, which means that the duration of PTE can be significantly reduced.

\subsection{Motivation}
Although the PHY and MAC layer of the PLC network have been studied and applied in various contexts, there is still a space for improvement when it comes to the design of a more efficient and robust network for IIoT. The designated bandwidth of IEEE1901.1 is greater than the available bandwidth of actual power line channel in industrial settings, and it uses long data frames, which makes the network operate slowly. In practical situations, it takes more than ten minutes to connect over 100 nodes. As for the P-MAC mechanism, it possesses reduced networking time, but further research is needed to ensure its robustness and scalability in larger networks. Hence, beyond the existing protocols such as IEEE1901.1 and P-MAC improved MAC protocols should be designed to tackle narrow bandwidth and large number of nodes in industrial settings.
Therefore, in this paper, we design a new robust preamble based MAC mechanism in PLC based IIoT, as detailed in the following text.

\section{Basic mechanism of R-PMAC}

In this section, we introduce the three basic techniques in R-PMAC, including the precise time difference calculation, preamble generation, and SID allocation. These detailed technical are the foundation of the R-PMAC mechanism.

\subsection{Precise Time Difference Calculation}
In the PTE stage of R-PMAC, CCO transmits a downlink preamble to STAs, and then STAs respond the CCO with uplink preambles. The time difference between the CCO's downlink preamble and the STAs' uplink preambles can help to distinguish different STAs. However, due to limited processing capabilities of the CCO and STAs, they are unable to handle signals in real-time, leading to uneven time differences measured made by the CCO and STAs.

\begin{figure}[htbp]
\flushleft
\includegraphics[width=8.5cm]{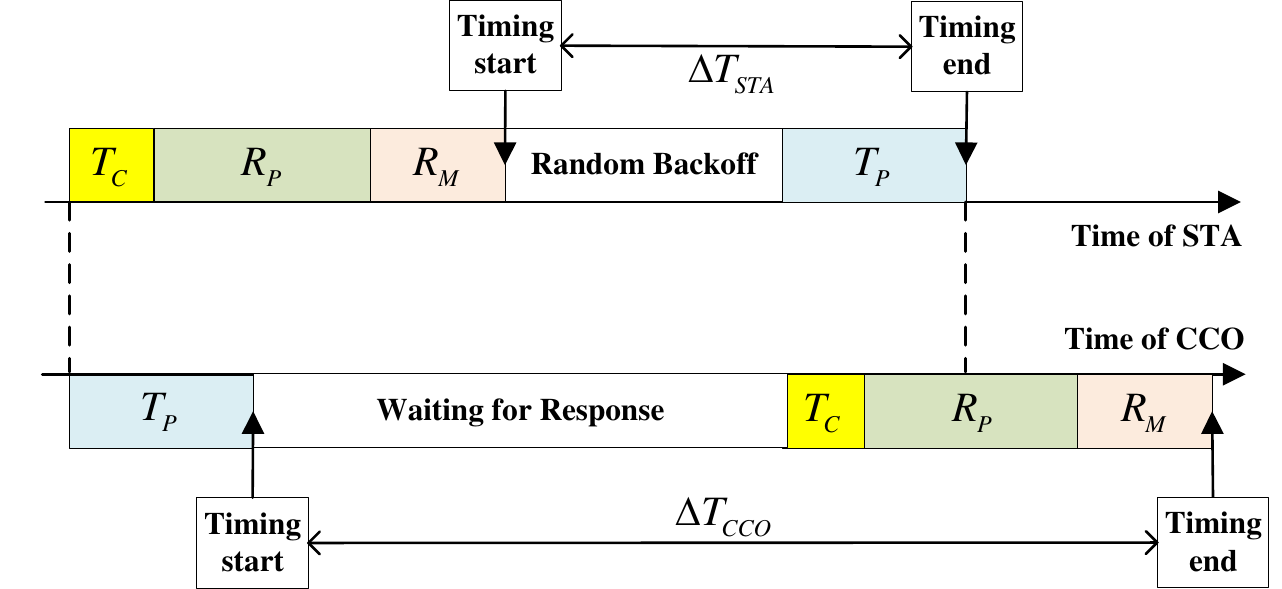}
\caption{Delay in PTE process. The CCO or STA starts timing after the received signal is sent to MAC layer, and ends timing after the signal to transmit is sent to PHY layer. The delays lengthen the time difference of CCO.}
\label{fig:deltaT}
\end{figure}

There are three main types of delays that contribute to the measurement errors: sending delay, propagation delay, and receiving delay. The receiving delay can be further divided into the time spent on signal processing by the PHY hardware and the data transportation between PHY hardware and MAC hardware. These delays are listed in TABLE \ref{tab:delay}, and Fig. \ref{fig:deltaT} illustrates how these delays cause the time differences of STA and CCO different. In IIoT, the distance between nodes is usually small, so the influence of propagation delay is much less than that caused by other delays. Therefore, $T_{C}$ can be ignored, while other delays in TABLE \ref{tab:delay} should be taken into account.
Therefore, implementing P-MAC requires sufficient processing speed to meet the premise of $T_P=R_P+R_M$, which can be challenging for the low-cost hardware.

\begin{table}[htbp]
\caption{Hardware Delay in preamble based MAC}
\begin{center}
\begin{tabular}{|c|c|c|}
\hline
\textbf{Notation}&\textbf{Types of Delay}\\
\hline
$T_{C}$&Signal propagation in channel\\
\hline
$T_{P}$&Coding and transmitting data to channel\\
\hline
$R_{P}$&Receiving and decoding data from channel\\
\hline
$R_{M}$&Sending data from PHY to MAC\\
\hline
\end{tabular}
\label{tab:delay}
\end{center}
\end{table}

\begin{figure}[htbp]
\centering
\includegraphics[width=6.5cm]{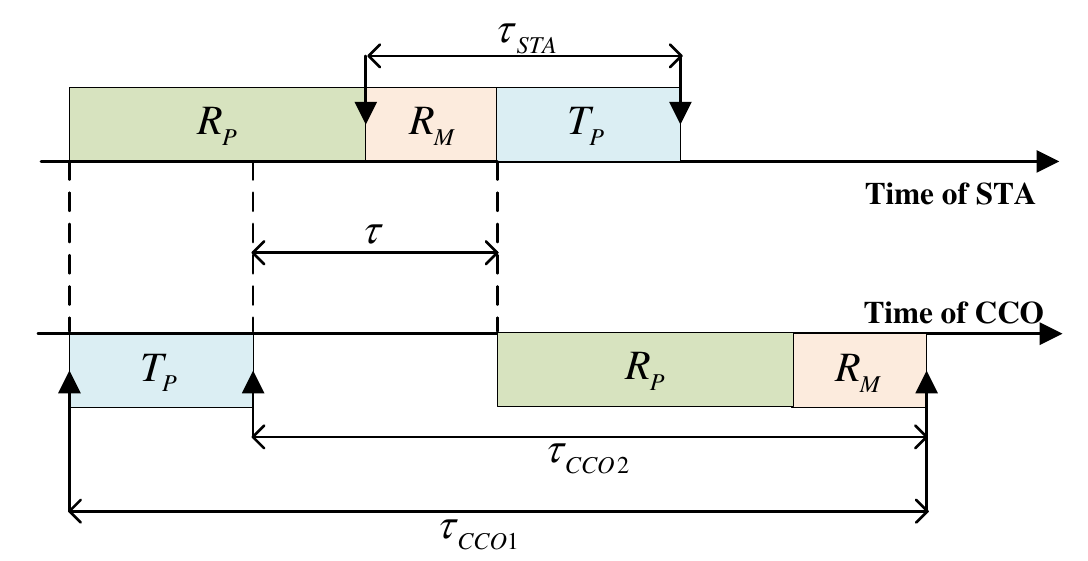}
\caption{Delay calibration mechanism of R-PMAC. In the process of delay calibration, the STA can be preset to respond to CCO without backoff, then the CCO can get the correction factor $\tau$ based on measured time difference.}
\label{fig:correction}
\end{figure}

Therefore, we introduce a precise time difference calculation mechanism to R-PMAC, allowing the CCO to obtain correct time differences even with a non-real-time receiver, making R-PMAC more robust than P-MAC. In OFDM-based systems, preambles and data frames have fixed lengths, so $T_P$, $R_P$, and $R_M$ in TABLE \ref{tab:delay} are constant. Based on this feature, we design the delay calibration mechanism shown in Fig. \ref{fig:correction}. The random backoff time of the STA can be set to zero. Then CCO measures the time differences $\tau_{CCO1}$ and $\tau_{CCO2}$, while the STA records $\tau_{STA}$ and conveys it to the CCO through a data frame. By introducing a correction factor $\tau$, linear equations about $T_P$, $R_P$, $R_M$, and $\tau$ are given. In practice, the $\tau_{CCO1}$, $\tau_{CCO2}$, and $\tau_{STA}$ can be easily measured in \eqref{eq:correction}, and the equations in \eqref{eq:correction} can be used to determine the values of the delays and the correction factor. For a specific hardware platform, the delays can be regarded as invariant, so the correction factor $\tau$ can be calculated and stored in the hardware as a constant. Then, in the PTE stage, the CCO can use \eqref{eq:corrected} to determine $\Delta\hat{T}_{STA}$, which is equal to $\Delta{T}_{STA}$.

\begin{equation}
\left\{
\begin{aligned}
    R_{P}+&R_{M}-T_{P}&-\tau &= 0\\
    R_{P}+&R_{M}      &+\tau &= \tau_{CCO2} \\
    R_{P}+&R_{M}+T_{P}&+\tau  &= \tau_{CCO1}\\
          &R_{M}+T_{P}&          &= \tau_{STA}
\end{aligned}
\right.
\label{eq:correction}
\end{equation}

\begin{equation}
    \Delta\hat{T}_{STA} = \Delta T_{CCO}-2\tau
\label{eq:corrected}
\end{equation}

\subsection{Preambles in R-PMAC}
\begin{figure} [t!]
	\centering
	\subfigure[\label{fig:REQ}REQ]{
		\includegraphics[scale=0.8]{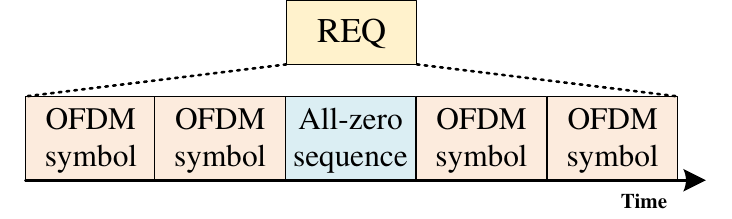}}
	\subfigure[\label{fig:NET}NET]{
		\includegraphics[scale=0.8]{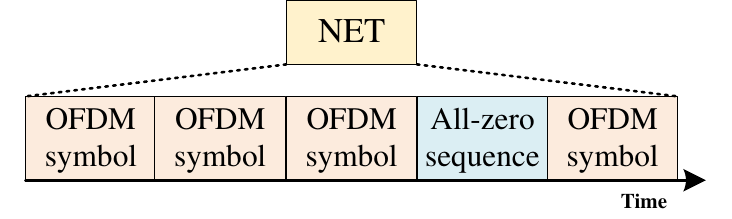}}
	\subfigure[\label{fig:DAT}DAT]{
		\includegraphics[scale=0.8]{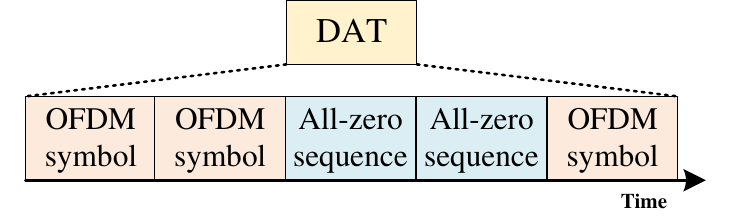}}
	\caption{The detailed structure of preambles. Every preamble contains five symbols with same length. The OFDM symbols has fixed waveform while the all-zero sequence is a blank interval}
	\label{fig:preambles} 
\end{figure}

In R-PMAC, preambles are used for not only frame synchronization but also time difference calculation. It is necessary to separate the preamble as three parts: the head of a data frame, the preamble transmitted by the CCO, and the preamble transmitted by the STA. Therefore, R-PMAC requires at least three different types of preambles and we refer to them as DAT (data frame preamble), NET (network preamble), and REQ (request preamble), respectively.
The method to generate these preambles has been discussed in \cite{PMAC}. Firstly, a short OFDM symbol is generated following the rule of least peak-to-average power ratio. Then, the preamble is formed by combining all-zeros sequences and the generated short OFDM symbol. The detailed structure of preambles is displayed in Fig. \ref{fig:preambles}, where the head and tail of the preamble are both marked with an OFDM symbol. The middle part of the preamble is formed by different combinations of all-zeros sequences and OFDM symbols to represent the type of preamble.

\subsection{SID allocation}
In order to properly manage unused SIDs and allot them to new STAs joining the network, CCO needs a suitable data structure to maintain them. There are several issues that need to be addressed when constructing the SID allocation mechanism.
\begin{itemize}
    \item Some SIDs might either fail to join the network or exit the network after a while. As a result, the previously used SIDs will have gaps with others as they are re-assigned to different STAs. Therefore, the CCO can't simply allocate the SIDs incrementally.
    \item The numerous SIDs contained in unproper data structures can easily get the CCO's memory exhausted. Thus, it is essential that the data structure be efficient in terms of storage.
    \item Given the large number of STAs, it is imperative that the CCO swiftly queries and updates the data structure for any unused SIDs.
\end{itemize}

Considering the second issue, a bitmap-style data structure is inappropriate. The third issue indicates that simple data structures like arrays and linked lists are inadequate for searching and updating SIDs because of their high time complexity. Here, we model our SID allocation mechanism after the operating system's memory allocation algorithm and implement it using a modified linked list.

\begin{figure}[htbp]
\centering
\includegraphics[width=8cm]{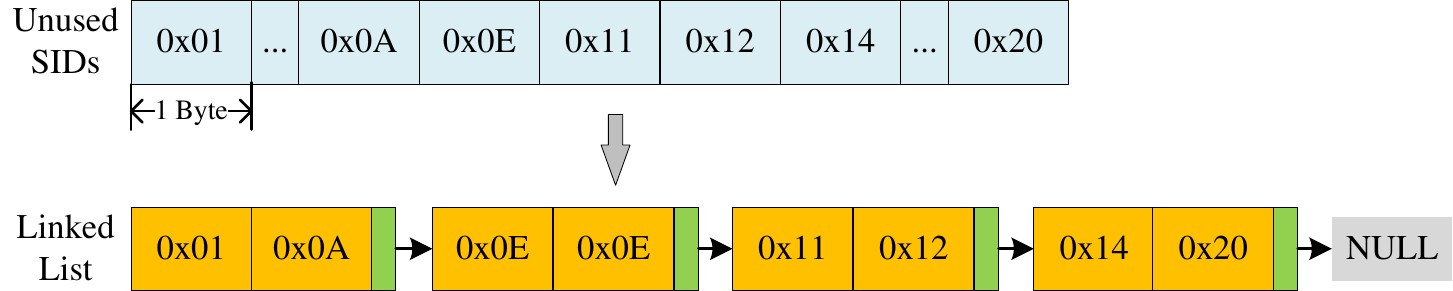}
\caption{An example of linked list generation. The contiguous SIDs from 0x01 to 0x0A are represented by a node while 0x0E, the dicrete SID is stored in aother node.}
\label{fig:linkedList}
\end{figure}

In the modified linked list, each node stores a segment of continuous SIDs represented with a minimum and a maximum SID. The Fig. \ref{fig:linkedList} shows an example of saving some discontinuous SIDs in such a linked list.

The modified linked list used for SID allocation in R-PMAC has a complexity of $O(1)$ for getting an idle SID or updating the data structure. The complexity of adding one SID is $O(M)$, where $M$ is the number of elements in the linked list. If all STAs successfully join the network, the idle SIDs are continuous and can be represented by a single node, saving space. In the worst case, every two idle SIDs are separated by one used SID, resulting in the largest possible size of the linked list, which is $O(N)$. Overall, this data structure has better performance and is more suitable for this scenario than bitmap, ordinary arrays or linked lists.

\section{Networking Process of R-PMAC}
PTE, T-Query, and Net-Config are the fundamental phases of a preamble-based MAC system \cite{PMAC}. In R-PMAC, these phases are organized into a networking cycle (NC) that occurs periodically. The networking process of a NC, e.g. the detailed scheduling of data frame transmission, needs to be carefully designed to address the following two key issues:
\begin{itemize}
    \item In an IIoT system with a large number of STAs, the networking process may spend substantial time because of inefficient scheduling of data frame transmission. This can slow down networking and make the PLC network sluggish. Therefore, the MAC layer protocol should establish connections between CCO and STAs using as few data frames as possible.
    \item Security-conscious networking procedure is necessary in some scenario to ascertain that only STAs with legal MAC addresses are allowed access. Accordingly, R-PMAC needs to have an authentication procedure, which means that all the MAC addresses need to be forwarded to R-PMAC through multi-hop.
\end{itemize}

The networking procedure of the tree-like PLC network can be subdivided into single-layer networking and multi-layer networking. Specifically, the single-layer networking elucidates the process of STAs connecting with CCO or PCO with one hop, while multi-layer networking defines how the CCO communicates with PCOs through multiple hops to expand the network. To resolving the aforementioned issues, we develop the single-layer and multi-layer networking procedure for R-PMAC.

\subsection{Single-layer Networking of R-PMAC}
\subsubsection{PTE stage}
The implement of PTE stage is relatively concise due to the attribute of preamble detection. Since the preamble carries no data and has limited duration, CCO and STAs cannot detect it until the transmission of preamble is finished. Therefore, complex carrier sensing is unnecessary in PTE. Once the CCO broadcasts the NET preamble, STAs wait for a random number of time slots and send REQ preambles. The random number ranges from 1 to a preset maximum number of time slots, recorded as $N_{max}$. The duration of time slots in PTE is very short (approximately 500$\mu$s), so the $N_{max}$ has little impact on the networking time. A maximum number of 256 slots has been considered sufficient in most situations.

\subsubsection{T-Query stage}
The data frames in T-Query are used to transmit time differences and MAC addresses. At the beginning of T-Query, the CCO broadcasts a data frame called Time Difference Frame (TDF), which contains the time differences recorded in PTE. Then the STA replies CCO with a data frame called MAC Address Frame (MAF), which contains its MAC address.

Using a TDMA approach to arrange MAF transmission is a viable option since it is simple to implement and can significantly save time spent in transmitting. The STAs iterate through the timing differences in the TDF. As soon as a STA detects a time difference in TDF consistent with the one it has recorded, it will get the index of that time difference, which we label as $i$. The CCO begins waiting for the MAF from the STAs after sending the TDF, and the STA transmits its MAF at the $i$-th time slot. Data frames take up lengthy time windows, so the existence of idle time windows may significantly elongate the networking process. By the means mentioned above, there are no free spaces in the schedule when using the methods described above.

\subsubsection{Net-Config stage}
The Net-Config stage uses data frames to transmit MAC addresses, SIDs, and acknowledgement signal indicating that a STA has successfully joined the network. Net-Config works similarly to T-Query. At the start of Net-Config, the CCO broadcasts a data frame called the Short ID Frame (SDF), containing the MAC addresses of the STAs and the SIDs assigned to them. The STAs decode the SDF, search for their SIDs, and update their network condition. Finally, the STAs that successfully join the network send Acknowledge Frames (ACK) to the CCO to inform it that they are available.

Using TDMA to manage the transmission of ACKs is feasible as well. The STA may look up its MAC address's index in the SDF and then transmit its ACK in the corresponding time slot.

\begin{figure*} [ht]
\centering
\includegraphics[width=18cm]{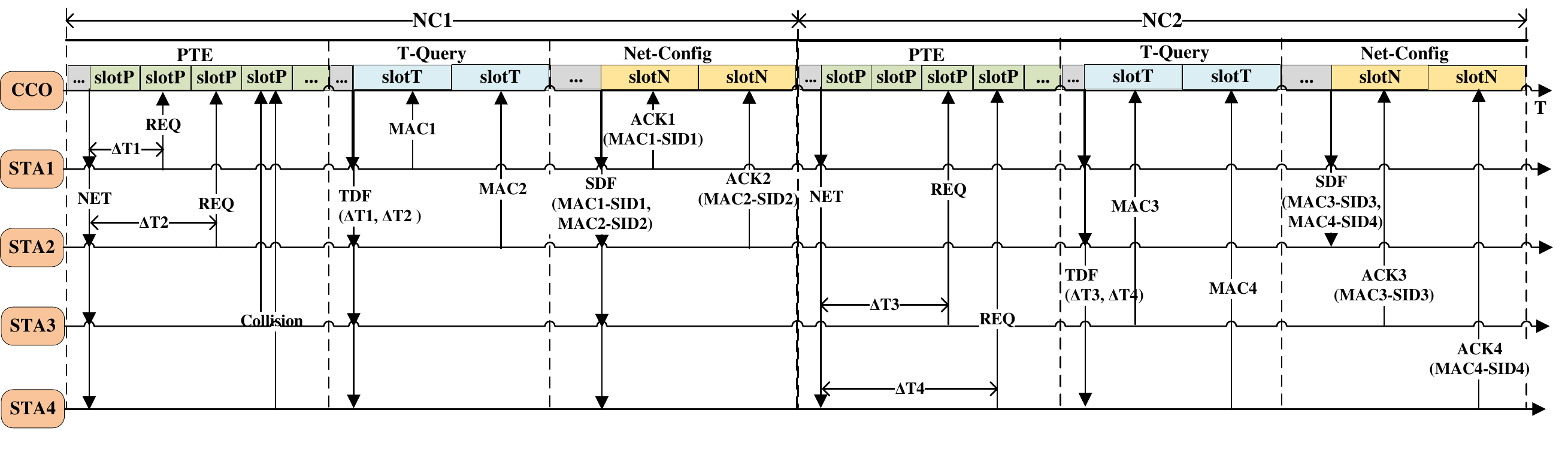}
\caption{Single-layer networking of R-PMAC}
\label{fig:RPMAC-single}
\end{figure*}

To demonstrate the R-PMAC single-layer networking method, we consider an example where a CCO connects with four STAs in two NCs. The process is depicted in Fig. \ref{fig:RPMAC-single} and consists of three steps.
First, the CCO in NC1 sends a NET preamble to signal the start of networking. The STAs then transmit REQs at random slots. STA1 and STA2 transmit REQs in different slots and are successfully heard by the CCO. The time difference can be represented by the number of time slots between NET and REQ. Considering STA3 and STA4, they send REQs in the same slot, which causes a collision. Consequently, the CCO does not detect legal preamble and disregards the preambles from STA3 and STA4.
Second, in T-Query, the CCO transmits a TDF containing the time differences of STA1 and STA2 ($\Delta T_1$ and $\Delta T_2$). As the time difference of STA1 in TDF is greater than that of STA2, STA1 transmits MAF in the first time slots while STA2 transmits in the second. STA3 and STA4 give up joining the network in this NC because they do not find their time differences in TDF.
Thirdly, in Net-Config, STA1 and STA2 receive their SIDs from SDF, update their online status, and send ACKs to the CCO. Therefore, STA1 and STA2 have successfully connected to the CCO in NC1. In NC2, the CCO sends another NET preamble in PTE while STA3 and STA4 send their REQs in different slots. Finally, STA3 and STA4 join the network with the same way to NC1.

\subsection{Multi-layer Networking of R-PMAC}
Due to limited communication distance of PLC, CCO needs some STAs to serve as PCOs and establish multi-hop links with farther STAs. In the meanwhile, considering the compulsory whitelist authorization, since only CCO has a whitelist, PCOs cannot directly allow STAs to join the network. Instead, it needs to promptly send all STA information to CCO. Hence, the interaction between CCO and PCOs in multi-layer networking should be appropriately designed.

\subsubsection{Behavior of PCOs}
We further divide PCOs into two types, Direct PCO (D-PCO) and Indirect PCO (I-PCO). D-PCO communicates with STAs directly while I-PCO serves as intermediate nodes between the CCO and the D-PCO. Specifically, D-PCO processes PTE, T-Query and Net-Config to connect with STAs. I-PCOs simply construct the route between CCO and D-PCO, and forward data frames. During multi-layer networking, the detailed route is determined and loaded to data frames by CCO. Then the I-PCOs can forward the data frames according to the route in data frames.

\subsubsection{Interaction between CCO and D-PCO}
The relationship between CCO and D-PCO is that CCO controls D-PCO to execute single-layer networking and find more STAs, while D-PCO feeds back these STAs' information to CCO. The data frames used by CCO and D-PCO are listed in TABLE \ref{tab:CCO-DPCO}.

\begin{table}[htbp]
\caption{Data frames between CCO and D-PCO}
\begin{center}
\begin{tabular}{|c|c|c|}
\hline
\textbf{Data frames}&\textbf{Functions}\\
\hline
\multirow{2}{*}{PTE-S}&CCO instructs an online STA\\& to be a D-PCO and start networking.\\
\hline
PTE-F&D-PCO sends back the number of STAs heard in PTE.\\
\hline
Tquery-S&CCO instructs D-PCO to start T-Query.\\
\hline
Tquery-F&D-PCO sends back the MAC addresses of the STAs.\\
\hline
Netconfig-S&CCO sends the SIDs allocated to the STAs.\\
\hline
Netconfig-F&D-PCO sends back the SIDs of online STAs.\\
\hline
\end{tabular}
\label{tab:CCO-DPCO}
\end{center}
\end{table}

\subsubsection{Detailed process of multi-layer networking}
The detailed process of multi-layer networking can be divided into 4 steps. Firstly, D-PCO receives the PTE-S freqme from CCO and starts the PTE stage to record some time differences. D-PCO then fills the PTE-F frame with the number of time differences and send it to CCO. Secondly, CCO transmits TQuery-S to D-PCO. After getting TQuery-S, D-PCO starts T-Query stage and gets the MAC addresses of the STAs which response to D-PCO in PTE stage. Thirdly, D-PCO conveys STA's MAC addresses to CCO through TQuery-F. CCO checks its whitelist, allocates SIDs to those legal MAC addresses, and delivers the SIDs to D-PCO through NetConfig-S. Considering that the TQuery-F and NetConfig-S have limited length and cannot accommodate the whole MAC addresses, the third part may repeat for some times until all the MAC addresses are checked by CCO. Fourthly, D-PCO gets all the SIDs from CCO, so it starts Net-Config stage and gets ACKs from STAs which successfully join the network. Then D-PCO transmits the SIDs of the newly joined STAs back to CCO through Netconfig-F.

\begin{figure*} [ht]
\centering
\includegraphics[width=18cm]{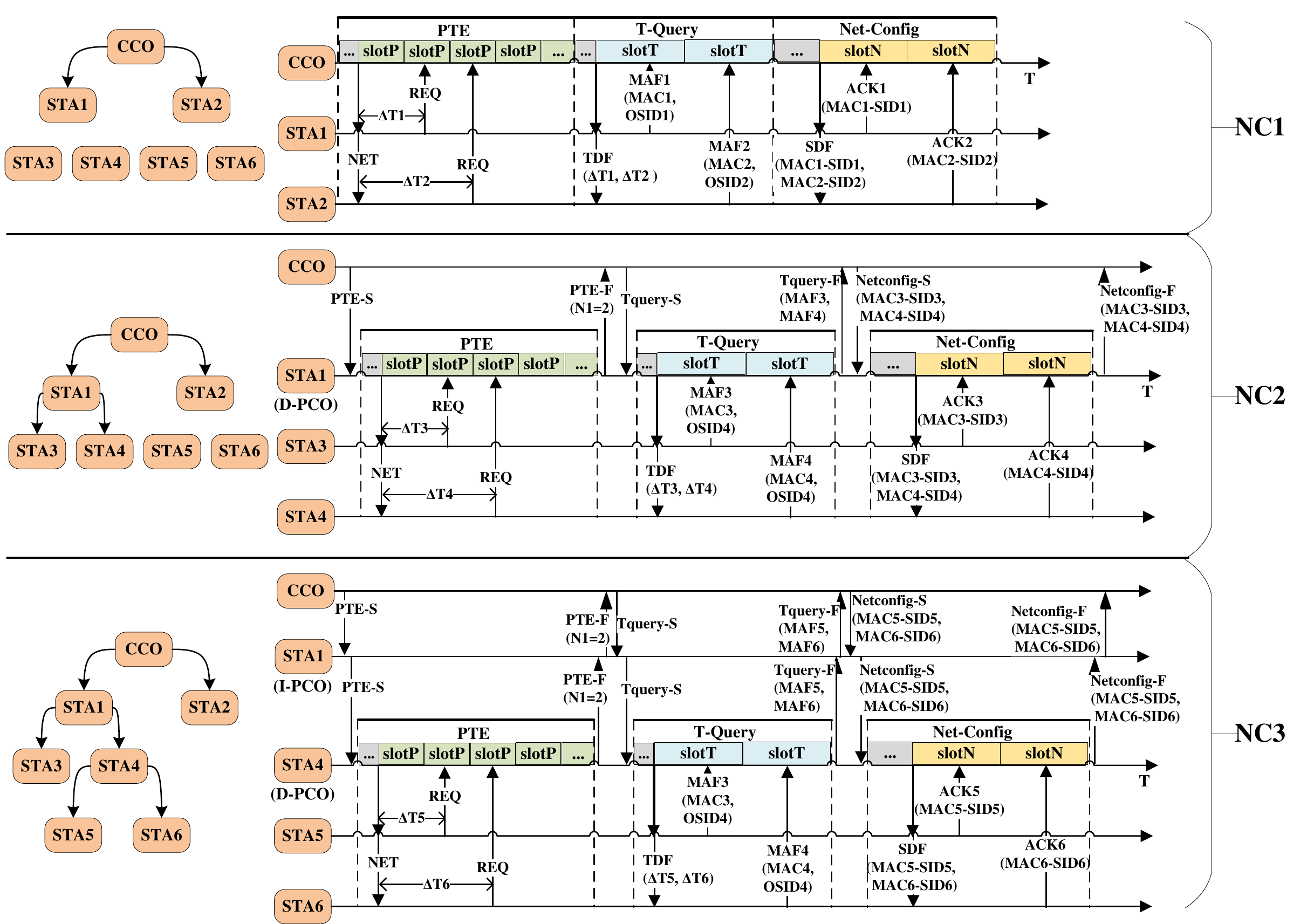}
\caption{Multi-layer networking of R-PMAC}
\label{fig:RPMAC-multi}
\end{figure*}

Let's consider an example scenario in which CCO needs to connect with six STAs across two layers, as shown in Fig.\ref{fig:RPMAC-multi}. In this scenario, the CCO can only directly communicate with STA1 and STA2. STA1 can connect with STA3 and STA4, while STA4 can connect with STA5 and STA6. In NC1, STA1 and STA2 connect to the CCO. In NC2, CCO sends a PTE-S to STA1 to instruct it to start the PTE stage. STA1 records the time differences of STA3 and STA4 and sends a PTE-F to CCO with the number of time differences (N1=2). CCO then sends a TQuery-S to STA1, making STA1 to initiate a T-Query stage and obtain the MAC addresses of STA3 and STA4. Then, STA1 forwards the MAC addresses to CCO through TQuery-F. CCO checks that these MAC addresses (MAC1 and MAC2) are in the whitelist and assigns SIDs to them. It then sends a Netconfig-S to STA1. With the assigned SIDs, STA1 starts Net-Config and distributes the SIDs to STA3 and STA4. Finally, STA1 obtains ACKs from STA3 and STA4 and sends a Netconfig-F to CCO, which updates the states of STA3 and STA4 to be online. In NC3, STA1 serves as an I-PCO to connect with STA4, which serves as a D-PCO to connect with STA5 and STA6. Following the same procedure above, CCO is able to admit STA5 and STA6 to the network via STA4.

\subsection{Mathematical analysis of different MAC mechanism}
As the number of nodes in IIoT grows, it is imperative to implement a MAC mechanism that spends as little time as possible on networking. In this section, we analyze the performance of R-PMAC by comparing it with existing MAC mechanisms, including P-MAC and the MAC layer of IEEE 1901.1.

\subsubsection{Basic idea of analysis}
We analyze the performance of these MAC mechanisms by considering two networking processes: random access and communication. The number of nodes, denoted by $N$, and the number of slots, denoted by $M$, are used to calculate the mathematical expectation of the number of slots required for successful random access, represented by the random variable $X$. This calculation allows us to estimate the networking time expectation for each MAC mechanism.

\subsubsection{Analysis of PTE stage}
The PTE stage is the random access stage of P-MAC and R-PMAC, where each node chooses one slot to send a preamble. Collisions can occur if multiple nodes choose the same slot. The successful transmission occurs only if the slot chosen by a node is not used by others.

\begin{figure*} [t!]
	\centering
	\subfigure[\label{fig:pte20}$N=20$]{
		\includegraphics[scale=0.38]{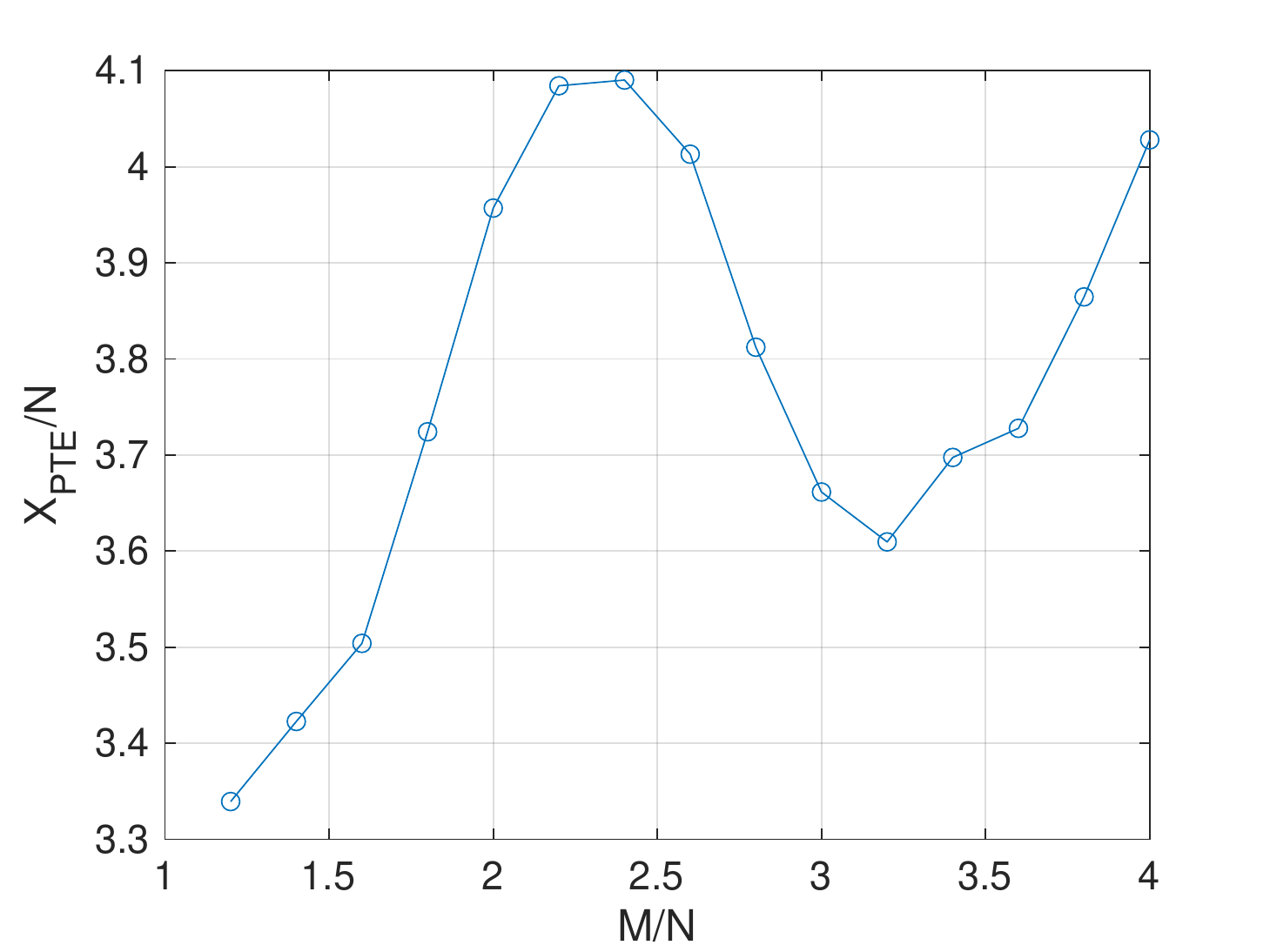}}
	\subfigure[\label{fig:pte100}$N=100$]{
		\includegraphics[scale=0.38]{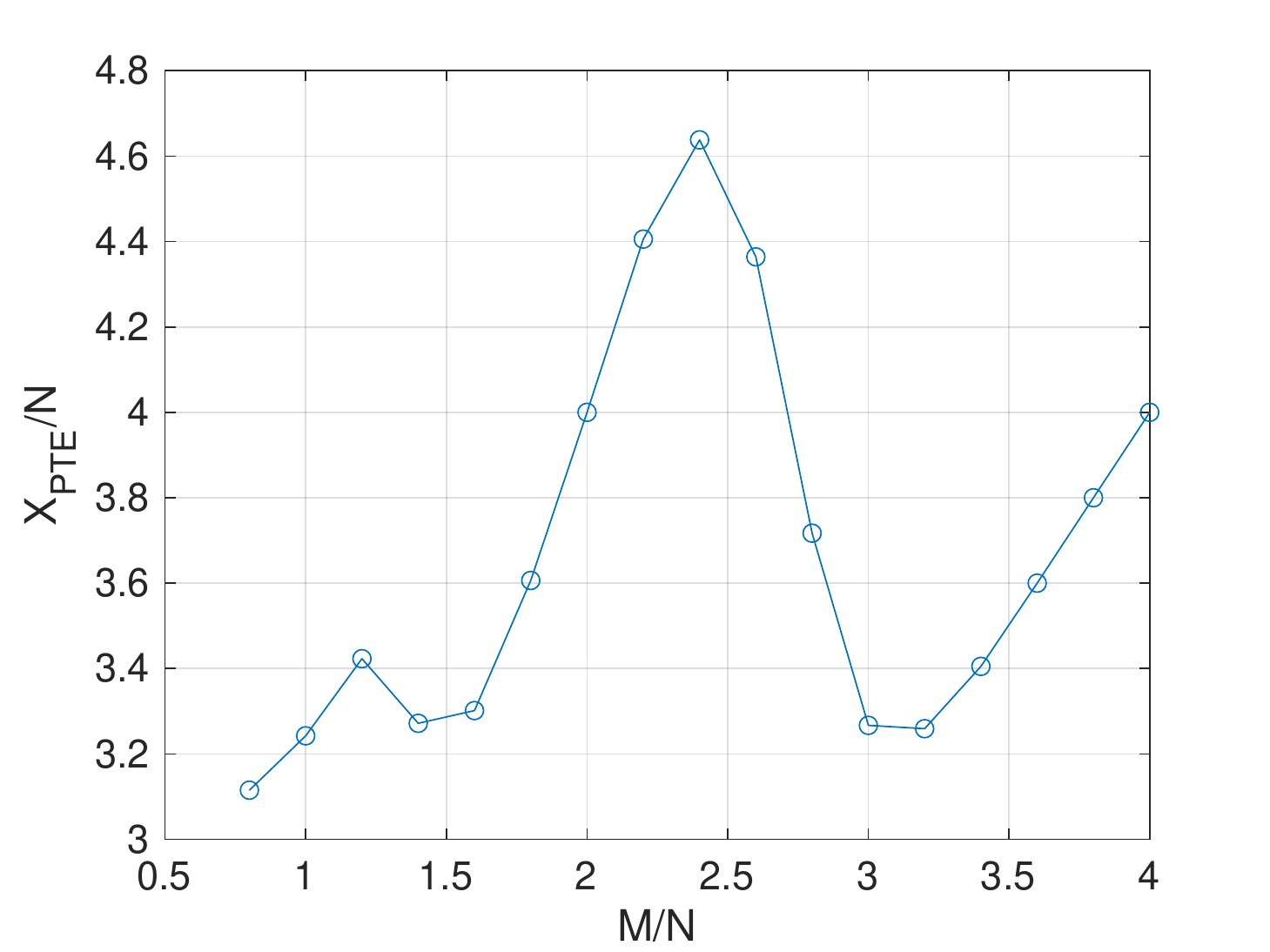}}
	\subfigure[\label{fig:pte200}$N=200$]{
		\includegraphics[scale=0.38]{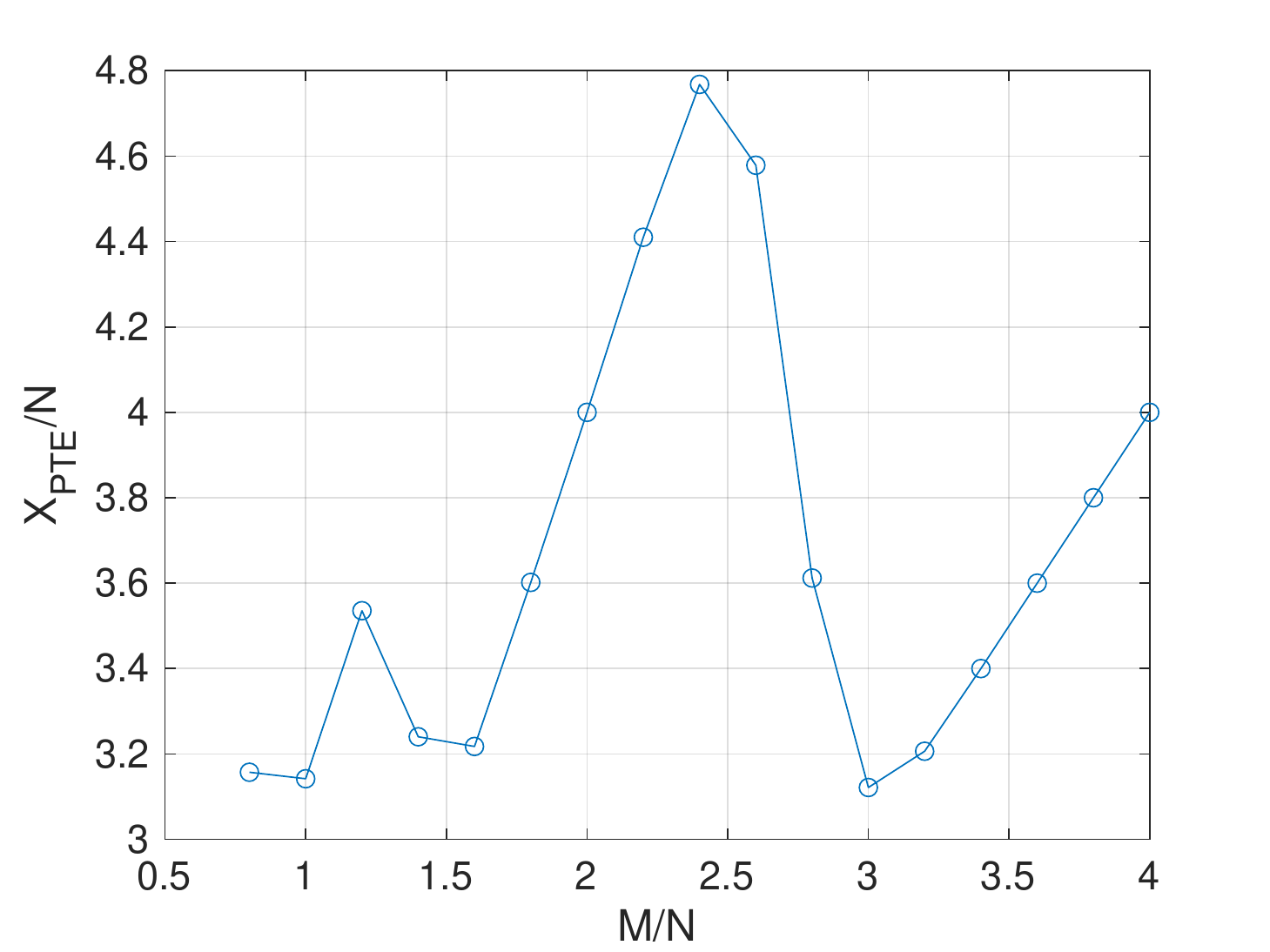}}
	\caption{Ratio of $X_{PTE}$ and $N$ versus ratio of $M$ and $N$. The simulation results shows that 4 is the integer closest to the values of $X_{PTE}/N$ in common cases.}
	\label{fig:pte}
\end{figure*}

To estimate the expected value of the random variable $X_{PTE}$ for PTE, we simulated the system with $N = {20,100,200}$ and $M/N$ ranging from 0.8 to 4. The simulation results in Fig. \ref{fig:pte} show that four PTE slots per node are required to access the CCO. Therefore, the expectation of $X$ for PTE can be represented by Eq. \eqref{eq:pte}.

\begin{equation}
E[X_{PTE}] = 4N
\label{eq:pte}
\end{equation}

\subsubsection{Analysis of CSMA stage}
The CSMA is the random access stage of the MAC layer of IEEE1901.1. To simplify the analysis, we assume that only one STA will be successfully heard by CCO if multiple STAs send data frames in one CSMA slot. Then the CSMA stage can be modelled as a Poisson process. We set the probability of one STA sending data frame in one CSMA slot to be $p$. Thus, the expectation of $X_{CSMA}$ can be given by \eqref{eq:csma}.

\begin{equation}
E[X_{CSMA}] = \frac{1}{p}N
\label{eq:csma}
\end{equation}

\subsubsection{Comparison of networking time}
We set the time length of PTE slot and CSMA slot to be $T_{PTE}$ and $T_{CSMA}$ respectively, time length of other data frames to be $T_{data}$, and $D(\cdot)$ to be the duration of networking time. Then the expectation of networking time with the three mechanisms is given by \ref{eq:networking}. The coefficient $1/p$ is close to 4. The $T_{PTE}$ is just the length of a preamble and the $T_{CSMA}$ is usually dozens of times longer than $T_{PTE}$. So we can conclude that preamble based MAC mechanism is faster than the CSMA mechanism used by IEEE1901.1. Compared with P-MAC, R-PMAC replaces polling with TDMA, which decreases the data frames sent by CCO and save 50 percents of networking time. In addition, R-PMAC decreases the data frames between CCO and D-PCO, which also saves time of multi-layer networking. So the advance of R-PMAC to P-MAC is more significant in multi-layer networking.

\begin{equation}
\begin{aligned}
    E[D(IEEE1901.1)]&=\frac{1}{p}NT_{CSMA}+4NT_{data}\\
    E[D(P-MAC)]&=4NT_{PTE}+4NT_{data}\\
    E[D(R-PMAC)]&=4NT_{PTE}+2NT_{data}
\end{aligned}
\label{eq:networking}
\end{equation}

\section{Robustness mechanisms in R-PMAC}
Unstable communication link and data frame collisions can occur during the networking process of a PLC network. The consequent the loss of critical data frames will interrupt the usual networking process. Ignoring these exceptions results in a lack of robustness and incurs malfunction in networking. Therefore, ensuring the robustness of the networking system is critical for our design. In this section, we examine the causes of data frame loss and propose solutions to address them.

\subsection{The consequences of data frame loss in PLC network}
There are two primary causes of data frame loss in PLC networks: hardware clock instability and time-varying channel quality.

\subsubsection{Unstable hardware clock}
As previously mentioned, the TDMA method replaces the polling method to reduce data frame transmission time and save networking resources. In TDMA, time slots are allocated to the STAs, which begin timing at the same time and send data frames at their assigned time slots. The TDMA method requires synchronized clocks for the STAs; otherwise, data frames from different STAs may overlap, resulting in data frame loss. Nevertheless, nodes in the PLC network typically have low-cost hardware with unstable crystal oscillators that fluctuate with time-varying temperature, preventing complete synchronization of the node clock. Due to this hardware flaw, the collision of data frames delivered by multiple STAs cannot be entirely eliminated, leading to data frame loss.

\subsubsection{Time-varying channel quality}
In addition to PLC nodes, varieties of appliances are also connected with the power line and generate severe current noise, which worsens the channel quality. In this situation, the route that has been established may be disconnected after a while. The data frames will be lost if they are transmitted through the disconnected route.

\subsection{The impact of data frame loss}
Data frame loss can have a negative impact on the preamble-based networking process. Loss of critical data frames can result in CCOs and STAs receiving incorrect information and taking incorrect actions, resulting in abnormal behavior. We will compare the consequences of data frame loss in the T-Query stage, Net-Config stage, and multi-layer networking.

\subsubsection{T-Query stage}
In T-Query stage, STAs send MAFs to the CCO with the method of TDMA, so collision and loss of MAFs due to unstable clocks may occur. And the TDF may loss due to poor channel quality. Loss of MAFs and TDF does not conveys wrong information but reduces the number of MAFs received by CCO, so CCO has to start more NCs to connect with the whole STAs.

\subsubsection{Net-Config stage}
Similar to the T-Query stage, STAs send ACKs to the CCO via TDMA, which can result in lost ACKs. Inadequate channel quality can also lead to SDF loss. Loss of ACKs and SDFs can result in an exceptional situation where the STAs and CCO inconsistently update the STAs' states. After receiving the SDF from the CCO, a STA will change its status to online, record the SID, and send an ACK to the CCO. If the CCO does not receive the STA's ACK, it will assume the STA is inaccessible, abandon attempts to establish a connection with the STA so that the SID can be recycled. In subsequent NCs, the same SID will be assigned to different STAs, resulting in two STAs with identical SIDs coexisting. They will simultaneously respond to the CCO, leading to a collision. Therefore, it is crucial to prevent Net-Config data frame loss.

\subsubsection{Multi-layer networking}
In multi-layer networking, data frame loss in T-Query and Net-Config still happens when D-PCO communicates with STAs. Besides, CCO interacts with D-PCO through the data frames listed in TABLE \ref{tab:CCO-DPCO}, which contain important information such as the MAC addresses of STAs and the SIDs of online STAs. Once one of these data frames is lost, CCO has to terminate the NC and all the STAs fail to join in the network in this NC. The CCO has to start more NCs to connect with the whole STAs and waste a lot of time.

\subsection{Mechanisms increasing robustness}
The issues of collision and poor channel quality need to be addressed in order to increase the robustness of PLC networks. While the probability of collision can be reduced by increasing the time slot length, excessively long time slots may slow down networking. Similarly, increasing the transmission power and improving signal-to-noise ratio can help reduce data frame loss, but this approach is not energy-efficient, which is contrary to the intention of IIoT. Hence, PLC networks require more sophisticated mechanisms to handle data frame loss and increase network robustness. In this paper, we propose a collision handling mechanism and a retransmission mechanism, which enables the network to function normally in the event of packet loss by adding small amount of extra data frames or attaching small amount of additional information to existing data frames.

\subsection{Collision handling mechanism}
Our analysis have shown that only the collision in Net-Config leads to logic error while that in T-Query only merely decreases the number of online STAs in one NC. Hence, we only try to deal with the collision in Net-Config.

\subsubsection{Basic idea of collision handling mechanism}
The basic idea of our collision handling mechanism is to combine the TDMA method with polling to avoid data frame collisions with the high time utilization of TDMA retained. Specifically, STAs first reply to the CCO in the TDMA method, and the CCO can detect data frame collision. For STAs that encounter collision, the CCO retries to communicate with them in a polling way.

\subsubsection{Detection of collision}
In ideal situation, the number of MAFs in T-Query and the number of ACKs in Net-Config received by CCO receives should be equal. In practice, however, the number of STAs in the Net-Config may be reduced due to the loss of ACKs. To detect collisions, the decrease in the number of STAs heard in Net-Config compared to T-Query can be used as an indicator of collision occurrence.

\subsubsection{Process of collision handling}
The Net-Config step is followed by the Polling phase. The CCO records the MAC addresses of the STAs that were lost in Net-Config and subsequently poll these STAs to establish communication. During the polling phase, the CCO may get ACKs from the recorded STAs, indicating that these STAs have obtained their SIDs and are now online. Their MAC addresses will be added to the list of online STAs by CCO. As for certain STAs that do not respond to CCO during the Polling step, CCO regards the link to them too unstable to exchange data. After the Polling stage, the states of the STAs recorded by CCO will match with those recorded by STAs.

\begin{figure*} [ht]
\centering
\includegraphics[width=18cm]{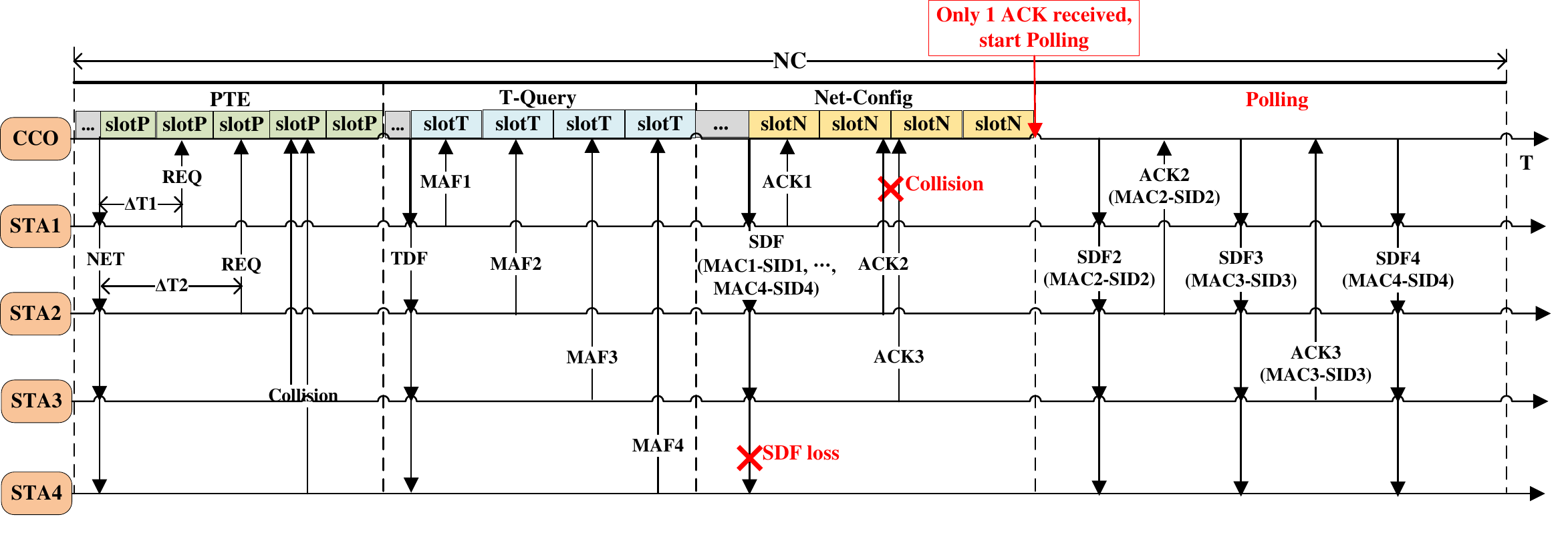}
\caption{An example of collision handling mechanism}
\label{fig:collision}
\end{figure*}

We provide an example to show the process of collision handling mechanism, which is shown in Fig.\ref{fig:collision}. The Polling stage follows the Net-Config, where the ACKs of STA2 and STA3 collide and STA4 dost not receive the SDF. The CCO detects collision when it only gets one ACK from STA1, and then starts the Polling stage, in which CCO sends SDF to STA2, STA3 and STA4 one by one and waits for their response. Since STA1 and STA2 have gotten their SIDs in Net-Config, they retransmit ACKs in Polling stage and get heard by CCO. As for STA4, it failed to receive its SID due to the lost SDF, so it does not reply to CCO. The CCO assures that STA4 cannot be online in this NC.

\begin{figure*} [ht]
\centering
\includegraphics[width=18cm]{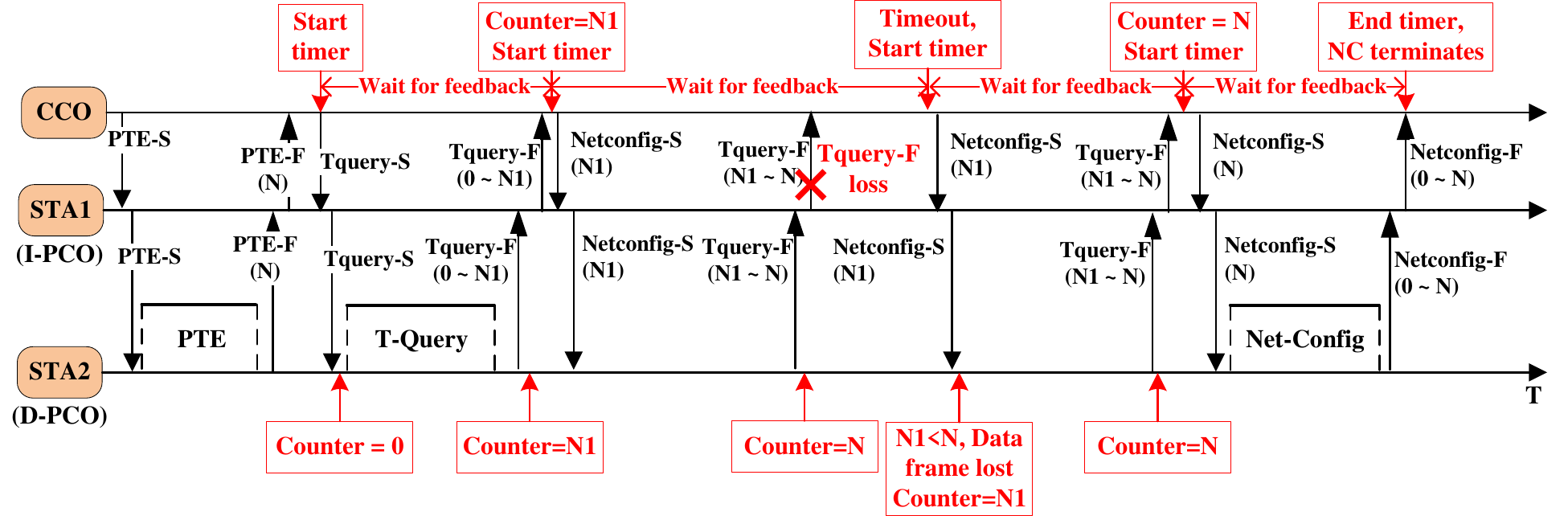}
\caption{An example of retransmission mechanism}
\label{fig:frame_loss}
\end{figure*}

\subsection{Retransmission mechanism}
\subsubsection{Basic idea of retransmission mechanism}
In multi-layer networking, the probability of data frame loss increases with the number of hops between CCO and D-PCO. Without a handling mechanism, these data frame loss may interrupt the NC. To prevent this, we add extra information to the data frames between CCO and D-PCO, allowing CCO and PCO to backtrack to their previous state and recreate the lost data frames.

\subsubsection{Detection of data frame loss}
As CCO knows the route from itself to D-PCO and the duration of a data frame, it can use the hop number to calculate the maximum waiting time. After transmitting a data frame to D-PCO, CCO starts a timer and awaits response. If the timer exceeds the maximum waiting time, the data frame is considered lost and needs to be retransmitted by D-PCO.

\subsubsection{Process of retransmission}
When D-PCO detects a retransmitted data frame from CCO, it uses the saved information to go back to the previous transmission state and reproduce the data frame sent last time. Moreover, if the timeout happens several times, the connection between the CCO and the D-PCO will be deemed unreliable, and the CCO will cease communicating with this D-PCO and control another one for subsequent networking.

An example is provided to illustrate the process of retransmission mechanism in Fig. \ref{fig:frame_loss}. As the number of STAs is large, D-PCO has to send MAC addresses in two TQuery-F. After receiving the first TQuery-F, CCO saves $N_1$, the number of STAs in the first TQuery-F. CCO then requests the second part of MAC addresses from D-PCO, but the second TQuery-F is lost during forwarding. CCO detects the data frame loss through its timeout timer and loads $N_1$ into a new Netconfig-S, which is sent to D-PCO. D-PCO checks the Netconfig-S and finds that $N_1$ is less than $N$, the number of STAs it sent to CCO, indicating that the last TQuery-F was not received by CCO. Hence, D-PCO loads the MAC addresses of the STAs ranging from $N_1$ to $N$ into a new TQuery-F and sends it to CCO. CCO receives the new TQuery-F successfully and gets the second part of MAC addresses. By this means, the multi-layer networking continues despite data frame loss.

\section{The format of data frames}
In addition to the robust networking process of R-PMAC, it is important to define the detailed format of data frames. In this section, we have designed the formats of various types of data frames, including their length and content.

\subsection{The length of data frames}
Each OFDM symbol contains the same amount of bits, and a fixed-length data frame can simplify hardware design. Therefore, we can set the length of all data frames to a reasonable fixed value. In the networking process, the transferred data types mainly include time difference, MAC address, SID, and other brief data. Additionally, after the network is created, the majority of sent data will usually consist of short-length sensor data and control signals. Thus, we fix the length of the data frame to 100 bytes, which can contain dozens of various types of short data and satisfy the needs of both the networking and operational phases of the PLC network.

\subsection{Basic framework of data frames}
\begin{figure}[htbp]
\centering
\includegraphics[width=8.0cm]{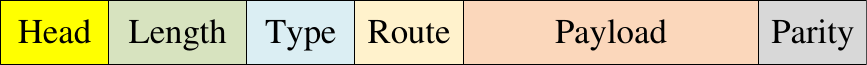}
\caption{Basic framework of data frames}
\label{fig:frame-basic}
\end{figure}

We clarify the structure of the data frame in Fig. \ref{fig:frame-basic}, which includes the head, valid data length, instruction type, route, payload, and tail. The head symbolizes the start of the data frame, and the valid data length indicates the length of the effective data in the frame. The instruction type is used to distinguish functions of data frames, and the route indicates the transmission path of the data frame. The payload is the actual data being transmitted, and the tail can be used for error parity to detect any errors in the data frame.

\begin{figure}[htbp]
\centering
\includegraphics[height=1.8cm]{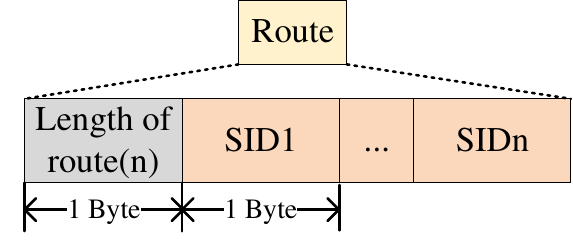}
\caption{The structure of route in data frame}
\label{fig:route}
\end{figure}

We use a one-byte data, 0x68, to represent the head, the detailed structure of the routing information is shown in Fig.\ref{fig:route}, in which the first byte represents the route length and the following bytes represent the SIDs of hops. We set the maximum number of nodes in the network to be 255, so the short addresses taken up 1 byte. In addition, since the maximum effective data length of the specified data frame is 100 bytes, the effective data length can be expressed with 1 byte.

One byte can be used to represent the instruction types, all of which are listed in TABLE \ref{tab:instruction}. Besides, this length is sufficient for adding more instruction types matching with other network functions such as data collection, channel quality detection, reboot of nodes, etc.

\begin{table}[htbp]
\caption{Instruction Types In Networking}
\begin{center}
\begin{tabular}{|c|c|c|}
\hline
\textbf{Instruction type}&\textbf{Hexadecimal value}\\
\hline
PTE&0x0B\\
\hline
T-Query&0x0C\\
\hline
Net-Config&0x0D\\
\hline
\end{tabular}
\label{tab:instruction}
\end{center}
\end{table}

\subsection{Payload formats of data frames}
Since the basic framework of the data frame has been determined, the payload format of data frames and the specific information can be clarified.

\subsubsection{Payload of TDF}
\begin{figure}[htbp]
\centering
\includegraphics[height=1.8cm]{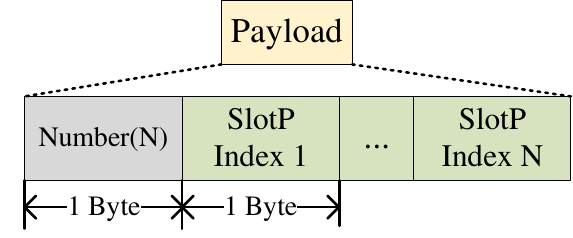}
\caption{The payload of TDF}
\label{fig:tdf}
\end{figure}

The payload of TDF is shown in Fig.\ref{fig:tdf}. It contains the number of time differences and indexes of slot in PTE, each of which takes up 1 byte. The time difference in practice usually is tens of thousands of microseconds and needs several bytes to save, which wastes the space of payload. Thus, we replace the time difference with the index of slot, which can be simply derived by dividing detailed time difference with time length of the slot in PTE. Since the maximum number of slots in PTE has been set to be 256, the index can be stored with only 1 byte.

\subsubsection{Payload of MAF}
\begin{figure}[htbp]
\centering
\includegraphics[height=1.8cm]{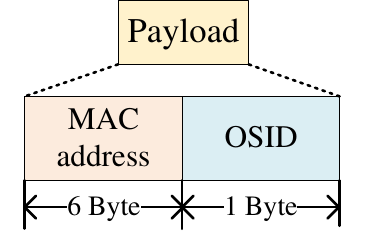}
\caption{The payload of MAF}
\label{fig:maf}
\end{figure}

The payload of MAF is shown in Fig.\ref{fig:maf}. It contains the MAC address and old SID(OSID) of the specific STA.

\subsubsection{Payload of SDF}
\begin{figure}[htbp]
\centering
\includegraphics[height=1.8cm]{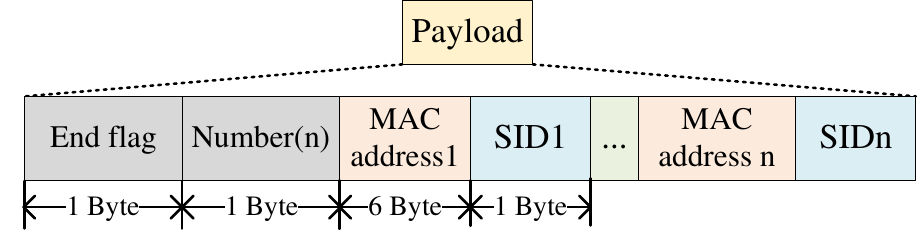}
\caption{The payload of SDF}
\label{fig:sdf}
\end{figure}

The payload of SDF is shown in Fig.\ref{fig:sdf}. It contains the number of MAC addresses and SIDs, and pairs of MAC addresses and SID. Besides, due to the limited capacity of one SDF, multiple SDFs may be sent successively and the End flag helps STAs to determine whether this SDF is the last one.

\subsubsection{Payload of ACK}
\begin{figure}[htbp]
\centering
\includegraphics[height=1.8cm]{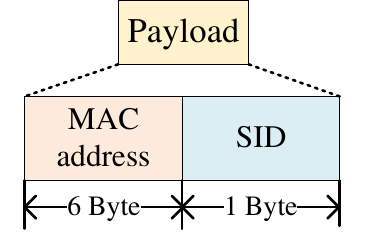}
\caption{The payload of ACK}
\label{fig:ack}
\end{figure}

The payload of ACK is shown in Fig.\ref{fig:ack}. It just contains its MAC address and updated SID.

\subsubsection{Payload of PTE-S, TQuery-S and PTE-F}
The payload of PTE-S, TQuery-S and PTE-F is simple. PTE-S and TQuery-S do not have payload because they just instruct the STA to start PTE and T-Query. The payload of PTE-F is just the number of STAs heard by D-PCO.

\subsubsection{Payload of TQuery-F}
\begin{figure}[htbp]
\centering
\includegraphics[height=1.8cm]{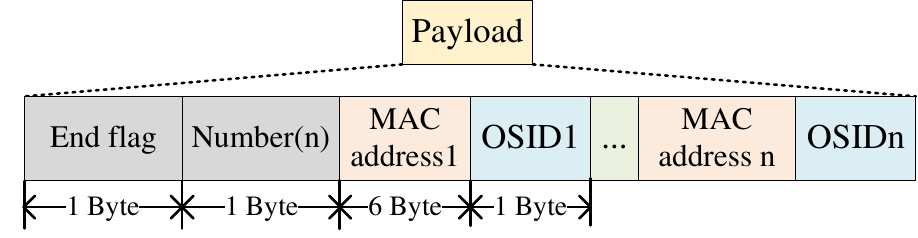}
\caption{The payload of TQuery-F}
\label{fig:TQuery-F}
\end{figure}

The payload of TQuery-F is shown in Fig.\ref{fig:TQuery-F}, which is similar to SDF. The distinction is that TQuery-F contains the OSIDs while SDF carries updated SID newly allocated by CCO.  When the D-PCO has sent all the MAC addresses of STAs, it sets the End flag to 1 to inform CCO.

\subsubsection{Payload of Netconfig-S}
\begin{figure}[htbp]
\centering
\includegraphics[height=1.8cm]{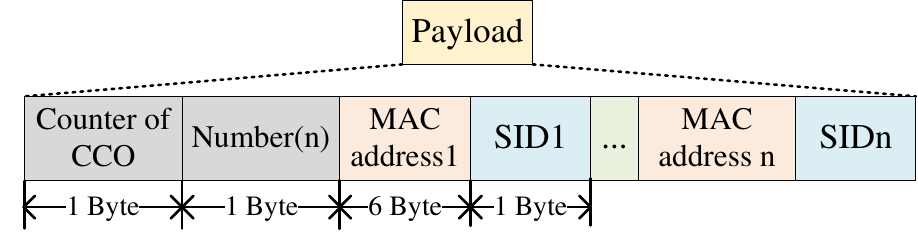}
\caption{The payload of Netconfig-S}
\label{fig:Netconfig-S}
\end{figure}

The payload of Netconfig-S is shown in Fig.\ref{fig:Netconfig-S}. Its structure is similar to the SDF. Specially, the counter of CCO records the number of STAs whose MAC addresses have been got by CCO. The counter of CCO is used in retransmission mechanism mentioned in $V.E$. If it is less than the counter of D-PCO, D-PCO will know that the last TQuery-F is lost.

\subsubsection{Payload of Netconfig-F}
\begin{figure}[htbp]
\centering
\includegraphics[height=1.8cm]{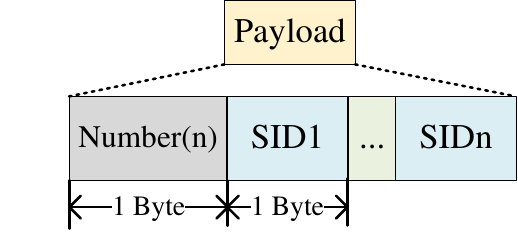}
\caption{The payload of Netconfig-F}
\label{fig:Netconfig-F}
\end{figure}

The payload of Netconfig-F is shown in Fig.\ref{fig:Netconfig-F}. It contains the SIDs of STAs joining in the network successfully.

\section{Design of Hardware Platform}
To validate the research presented in the previous chapters, we require a hardware platform that can run the networking process of R-PMAC, P-MAC, and IEEE1901.1 to compare their performance. In this chapter, we illustrate our design of the programmable hardware platform.

\subsection{Architecture of hardware platform}
When PLC networks work, the transmitting unit of hardware generates baseband signal based on the contents of the data packet and modulates it onto the power line. The receiving unit demodulates the signal from the power line and then does digital signal processing to retrieve the data frame. In addition, the networking operation must be controlled by a logical unit.

Analog front ends (AFE) are necessary to modulate and demodulate PLC signals and an embedded processor is needed to control the networking procedure and interact with the user. Coding and decoding baseband signals involve complicated mathematical calculations. Although the embedded processor may also operate as a calculating unit, simultaneously performing signal processing and networking process control will place a significant workload on it. As a result, we include FPGA, which can be configured as a digital front-end (DFE) for fast digital signal processing, serving as an interface between AFE and embedded processor. As for the analog signal, it is transferred through the power amplifier (PA) and coupler to the power line.

The FPGA and embedded CPU are programmable, thus we may implement various mechanisms on the same hardware platform. In addition, The circuit implemented in FPGA can be hardened as application-specific integrated circuits (ASIC) during afterward manufacturing process to save costs.

\subsection{Selection of hardware device}
To keep the cost low, we choose hardware devices that satisfy the basic requirement. We choose AD9865 \cite{AD9865} as the AFE. For embedded processor, we choose STM32F103RET6 \cite{STM32}, which is based on Cortex-M architecture and widely used by industry. For FPGA, we use Altera's Cyclone-IV chip \cite{CYCLONE}.

\subsection{Realization of hardware platform}
\subsubsection{The structure of hardware platform}
\begin{figure}[htbp]
\centering
\includegraphics[width=7.4cm]{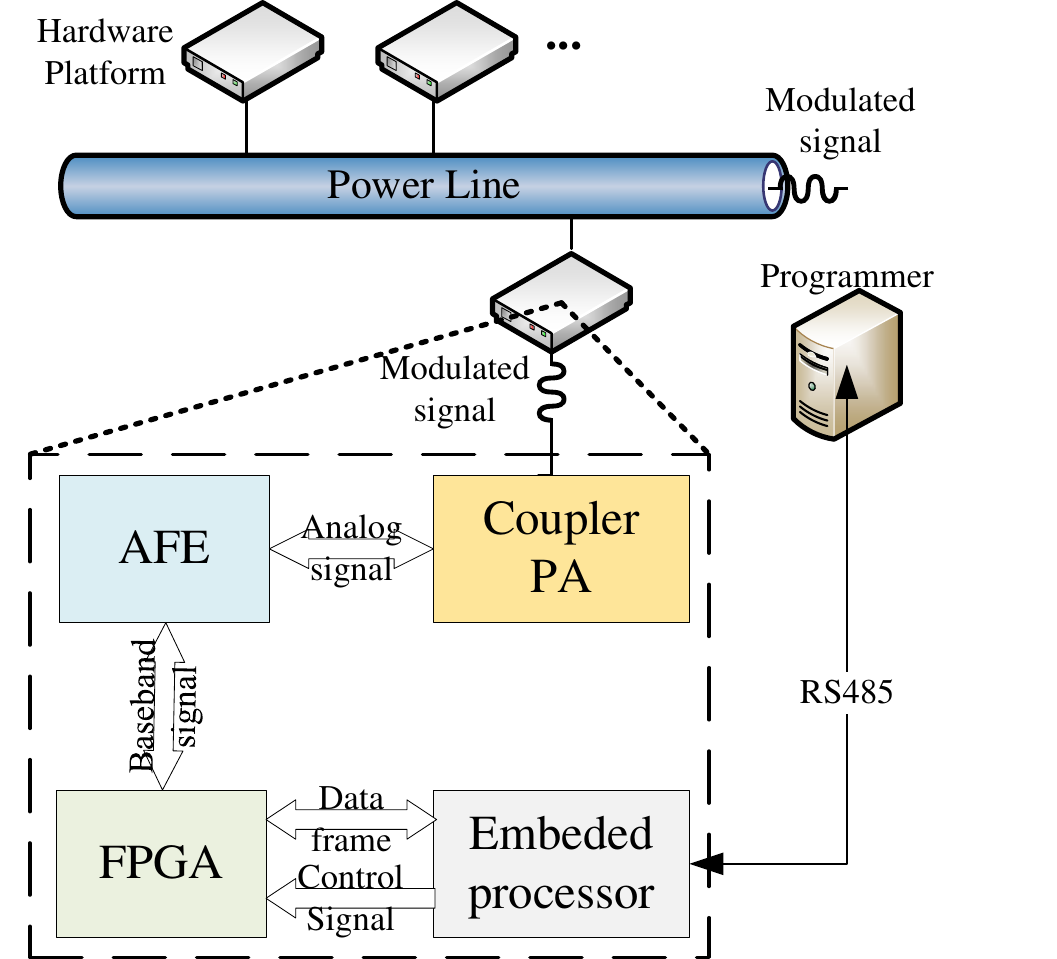}
\caption{The structure of hardware platform. The platform can be programmed with RS485 bus. The AFE does A/D and D/A converting. The data frames flow between the FPGA and the embedded processor, and the state of FPGA is configured by the embedded processor through SPI bus.}
\label{fig:hardware}
\end{figure}

The detailed structure of our hardware platform and its application is shown in Fig.\ref{fig:hardware}. There exist dozens or hundreds of nodes on the same power line. For one node, embedded processor control the FPGA and exchange data with FPGA. The RS-485 can be used to connect PC and the hardware platform.

\subsubsection{The digital front end in FPGA}
The DFE in FPGA is situated at the PHY layer. The detailed digital signal process done by the front end includes convolutional coding, Reed-Solomon (RS) coding, channel interleaving, and ROBO interleaving.

\begin{figure}[htbp]
\centering
\includegraphics[width=7.4cm]{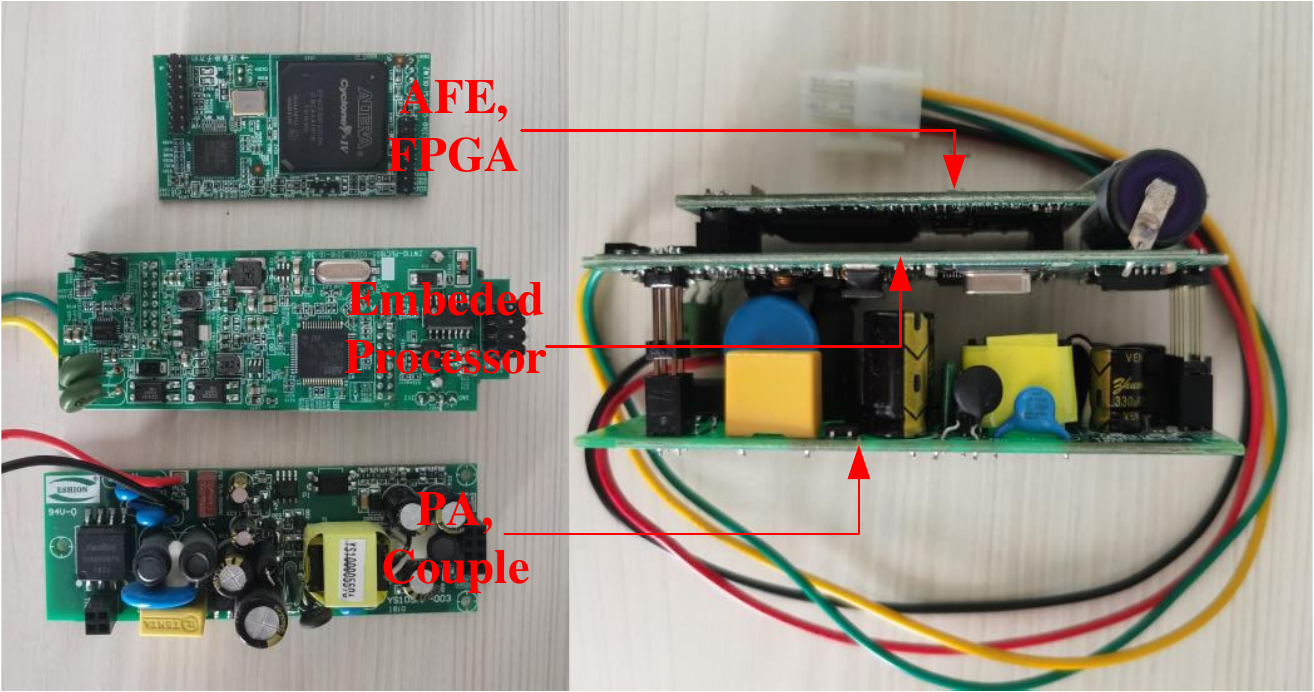}
\caption{The real hardware platform.}
\label{fig:node}
\end{figure}

Finally, the real hardware platform designed by us is shown in Fig.\ref{fig:node}. We partitioned the hardware platform into three boards, deploying AFE and FPGA, embedded processor, and power amplifier on each of them, respectively.

\section{Experiment}
In this chapter, we present our experiment and its results. We set the experiment parameters to match the physical environment and conduct measurements in a practical power line system to evaluate the performance of R-PMAC.

\subsection{Parameters of PHY Layer}\label{parameters}

We configure the parameters of hardware platform according to the parameters of IEEE1901.1. In this way, the result of hardware experiment can be closer to real situation and validate the superiority of R-PMAC to existing P-MAC and IEEE1901.1.

\subsubsection{Parameters of IEEE1901.1}
For the networking process of IEEE1901.1, data frames including central beacon, proxy beacon, association request message (MMeAssocReq) and association indication message (MMeAssocInd) are mainly involved \cite{IEEE1901}. Central beacon and proxy beacon are 127 bytes while MMeAssocReq and MMeAssocInd are 142 bytes. The beacons and messages are filled to 136 bytes and 272 bytes respectively and sent to PHY layer. Turbo coding, scrambling, channel interleaving, and ROBO (Robust OFDM) interleaving expand the bit length of beacons and messages to 21,760 and 43,520. The time length $T$ can be calculated by \eqref{eq:ieee1901}, in which $N_{s}$ is the number of OFDM symbol and $N_{c}$ is the number of feasible subcarriers. In actual industrial condition, available bandwidth is about 3MHz and $N_{c}$ is 94. Finally, the time length of beacons and messages are 9,102$\mu s$ and 17,488$\mu s$, respectively.

\begin{equation}
\begin{aligned}
    T=&40.96\times(13+N_{s})+18.32\times2\\
    &+(N_{s}-2)\times10.8(us),N_{s}=\lceil{N_{b}}/{N_{c}}\rceil
\end{aligned}
\label{eq:ieee1901}
\end{equation}

\subsubsection{Parameters of hardware platform}
Regarding the hardware platform parameters, we have previously established in Sections VI and VII that the length of the data frame is 100 bytes, and the FPGA processes convolutional coding, RS coding, channel interleaving, and ROBO interleaving. The convolutional and RS coding rates are set to 1/2 and 239/255, respectively, which increase the data frame to 8,768 bits. We use 1024-point OFDM symbols for data frames and 128-point OFDM symbols for the preamble. The bandwidth and sampling frequency are set to 1.25MHz and 40MHz, respectively, resulting in the preamble and data frames' time lengths becoming 51.2$\mu s$ and 819.2$\mu s$, respectively. For both P-MAC and R-PMAC, the preamble consists of 5 OFDM symbols, which takes 512$\mu s$. The available subcarriers number is 720, so the data frames with 8,768 bits take $512+8,768/720\times819.2=10,488\mu s$. We can see that the time length and data length of data frames used in the hardware platform are close to those in IEEE1901.1.

\subsubsection{Configuration of time slot length}
We set the time slot length of different data frames according to their time length. The detailed time length and time slot length of data frames and preambles are listed in TABLE \ref{tab:parameters}. Although the data frames in R-PMAC have shorter time length (10,488$\mu s$) than those in IEEE1901.1 (17,488$\mu s$), we set the time slot length in P-MAC and R-PMAC to be the same to that in IEEE1901.1 (20,000$\mu s$). Shorter networking time with identical slot length gets the superiority of R-PMAC more convincing.

\begin{table}[htbp]
\caption{Slot Time of Frames and Preamble}
\begin{center}
\begin{tabular}{|c|c|c|c|}
\hline
\textbf{Protocol}& \textbf{Frame Type}& \textbf{Time Length($\mu s$)}& \textbf{Slot Time($\mu s$)}\\
\hline
\multirow{4}{*}{IEEE1901.1}& Central Beacon& 9102& 12000\\
\cline{2-4}& Proxy Beacon & 9,102 & 12,000\\
\cline{2-4}& MMeAssocReq & 17,488 & 20,000\\
\cline{2-4}& MMeAssocInd & 17,488 & 20,000\\
\hline
\multirow{2}{*}{\begin{tabular}[c]{@{}l@{}}
R-PMAC/\\P-MAC\end{tabular}}& Preamble& 512& 600\\
\cline{2-4}& Data Frame & 10,488 & 20,000\\
\hline
\end{tabular}
\label{tab:parameters}
\end{center}
\end{table}


\subsection{Experimental results}
\begin{figure}[htbp]
\centering
\includegraphics[width=8.0cm]{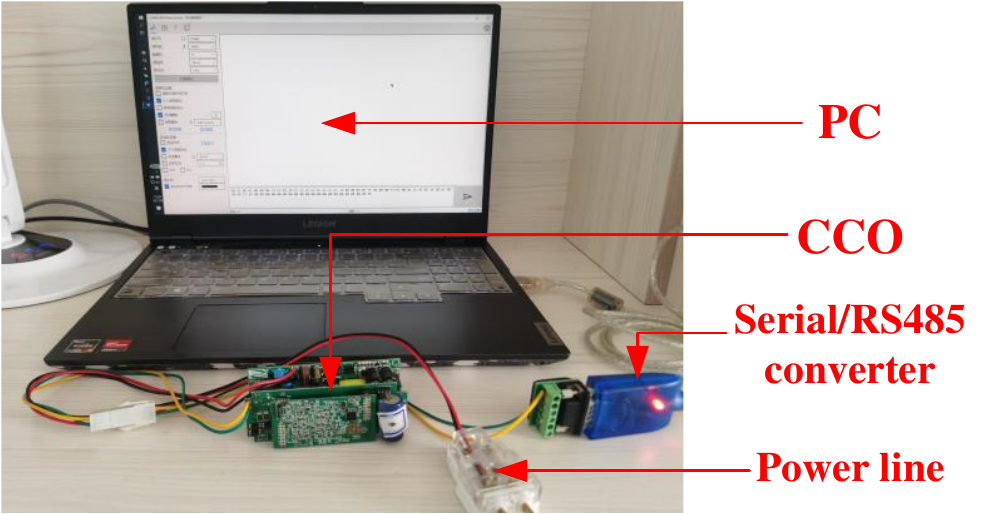}
\caption{The CCO connected with PC}
\label{fig:cco}
\end{figure}

We measured the networking time of a power line communication (PLC) network in a building with multiple residents and various appliances in use. The network consisted of CCO and several STAs connected to the building's power grid. The CCO was connected to a PC via a USB port for real-time data collection, as shown in Fig.\ref{fig:cco}.

We set the number of STAs to range from 90 to 240 and they can constitute a multi-layer network. Since the channel quality changes over time, the final network topology and the number of layers can vary randomly. In this situation, we recorded the average networking time for different numbers of layers. 

\begin{figure*}
	\centering
	\subfigure[\label{fig:layer3}Number of layers is 3]{
		\includegraphics[scale=0.48]{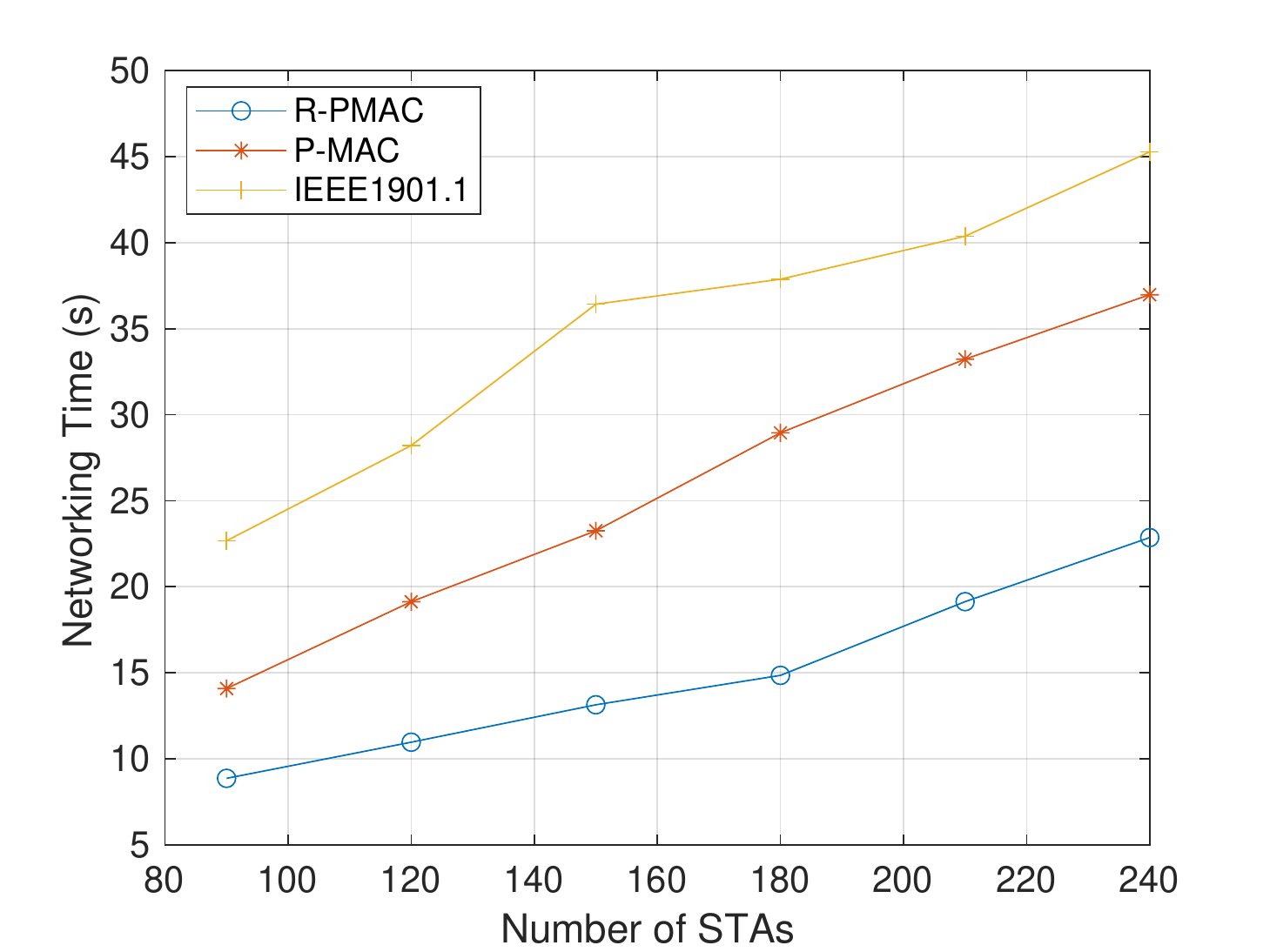}}
	\subfigure[\label{fig:layer4}Number of layers is 4]{
		\includegraphics[scale=0.48]{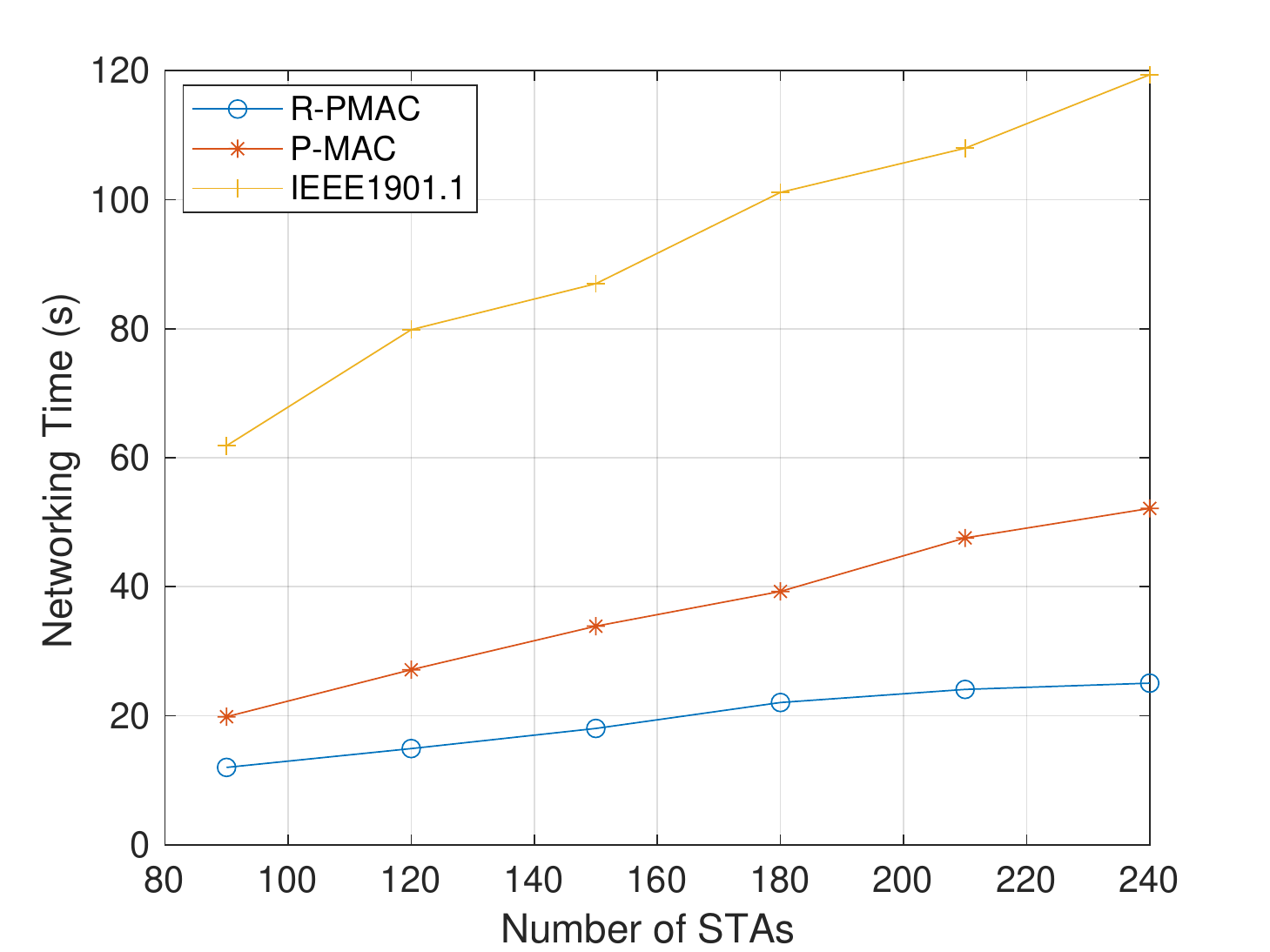}}
	
	\subfigure[\label{fig:layer5}Number of layers is 5]{
		\includegraphics[scale=0.48]{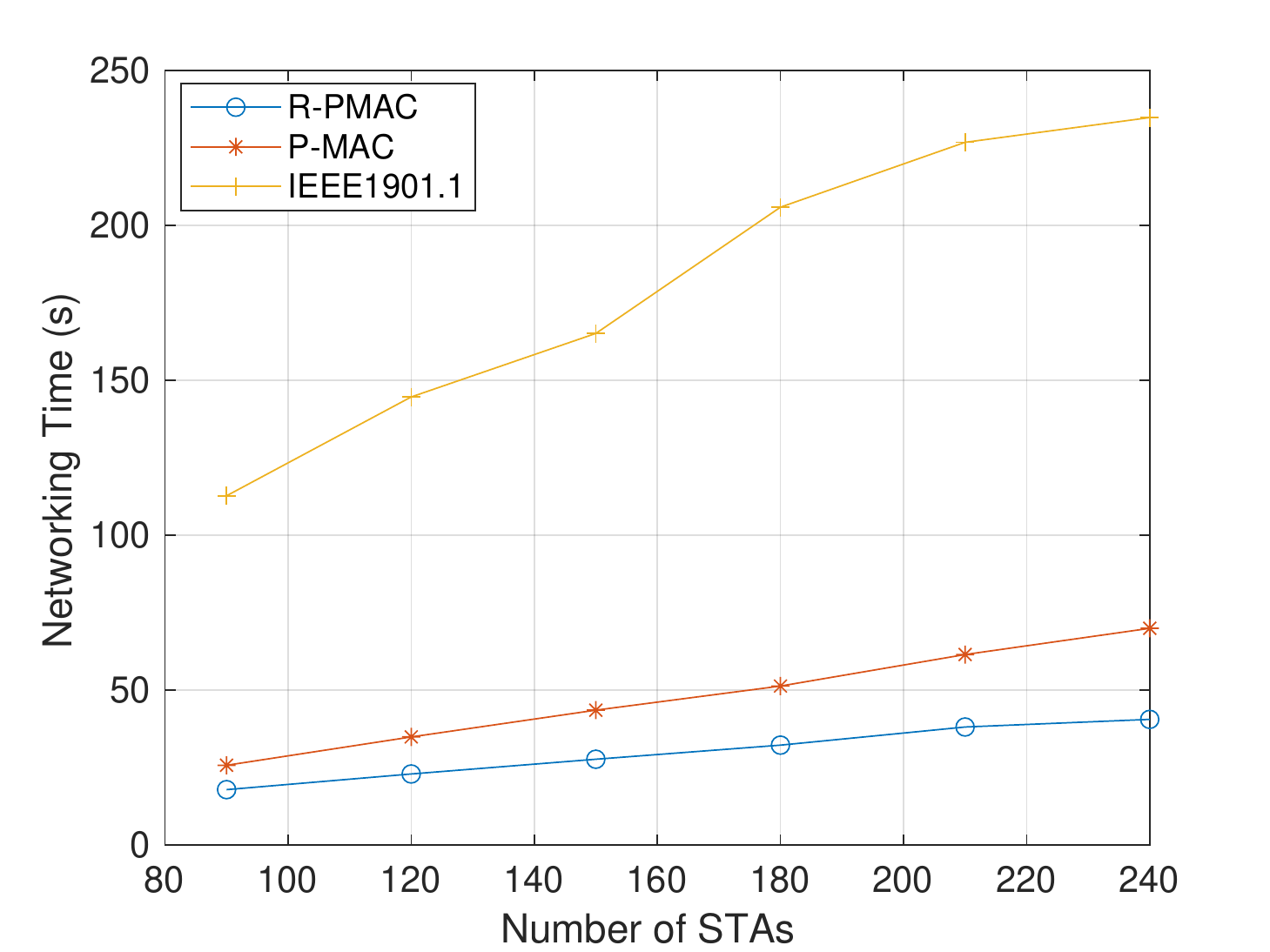}}
	\subfigure[\label{fig:layer6}Number of layers is 6]{
		\includegraphics[scale=0.48]{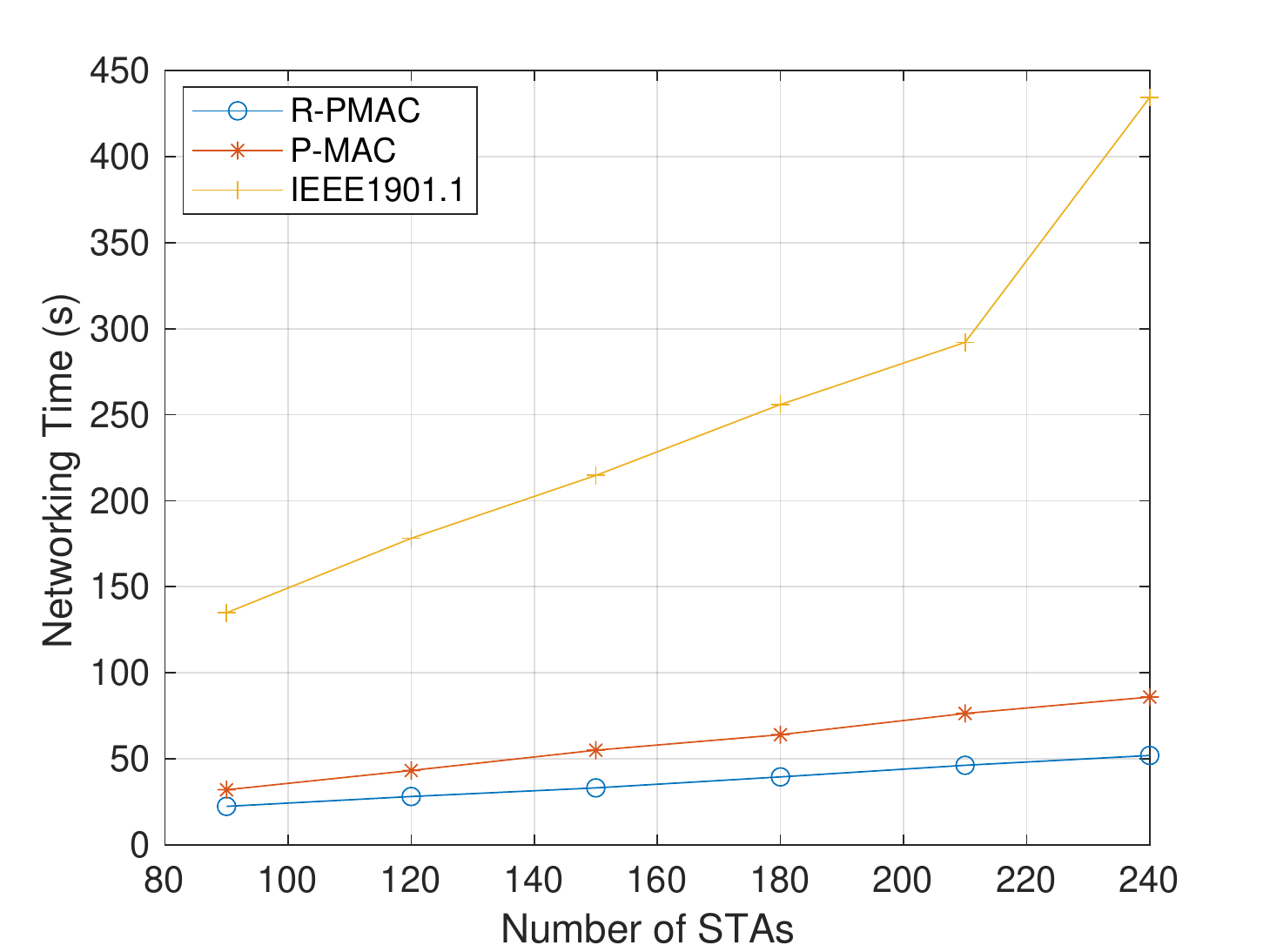}}
	\caption{Average networking time versus number of STAs for given number of layers. Given the number of STAs, the CCO will eventually construct a network with a certain number of layers, with which we classify the data and plot the average networking time under different number of STAs.}
	\label{fig:real}
\end{figure*}

The experiment result is shown in Fig.\ref{fig:real}. Based on the result, we find that:
\begin{itemize}
    \item The networking time with the three MAC mechanisms increase with the number of STAs and number of layers.
    \item The number of layers can have great influence on the networking time. The growth rate of networking time when using R-PMAC is significantly smaller than using P-MAC or IEEE1901.1.
\end{itemize}

We can draw the conclusion that R-PMAC can help the PLC nodes to organize a network at a higher speed. Besides, when using R-PMAC, the networking time of given number of STAs is not greatly affected by the maximum depth of network.

\begin{figure*}[t!]
	\centering
	\subfigure[\label{fig:without}Without robustness mechanism]{
		\includegraphics[scale=0.48]{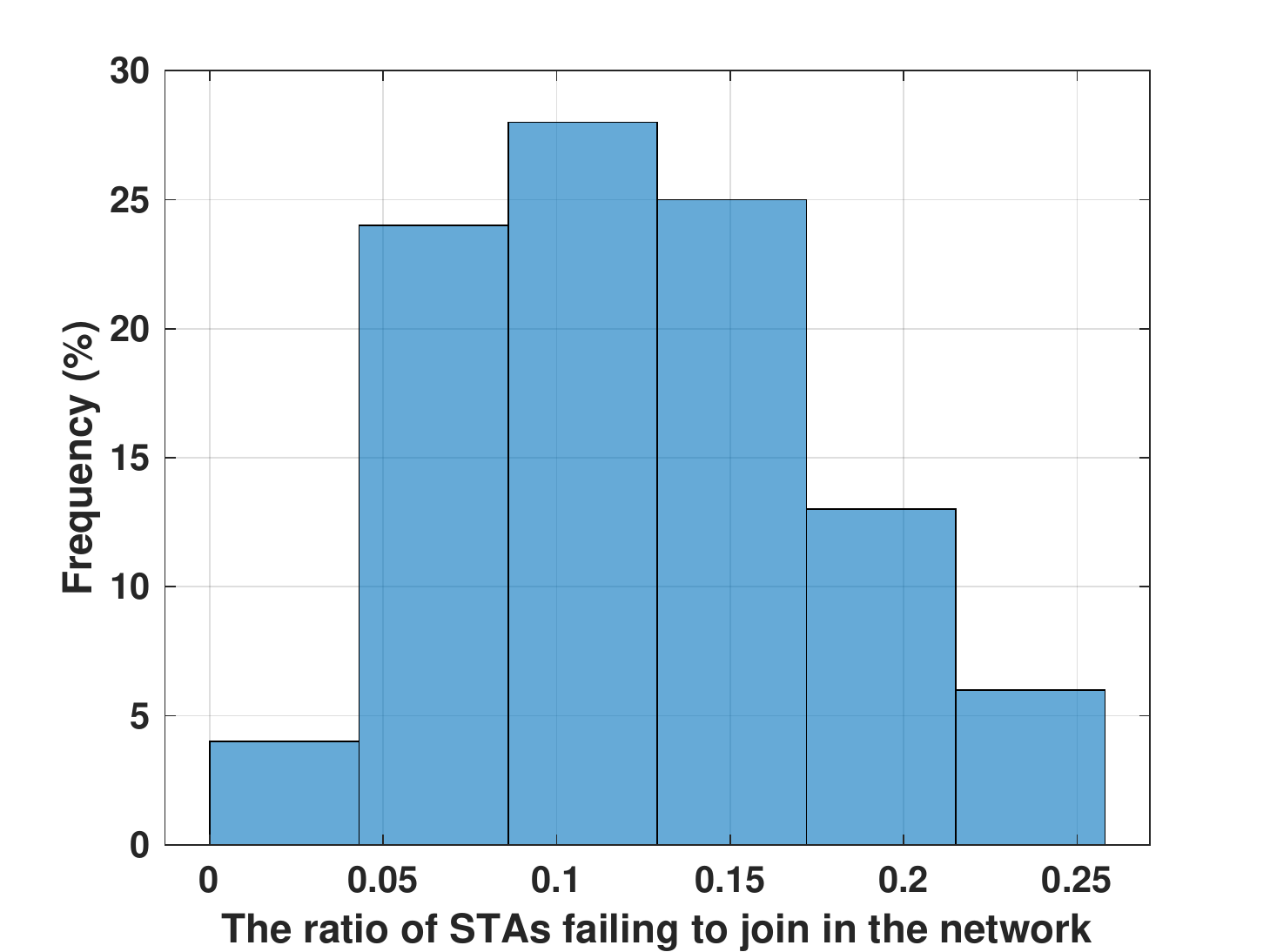}}
	\subfigure[\label{fig:with}With robustness mechanism]{
		\includegraphics[scale=0.48]{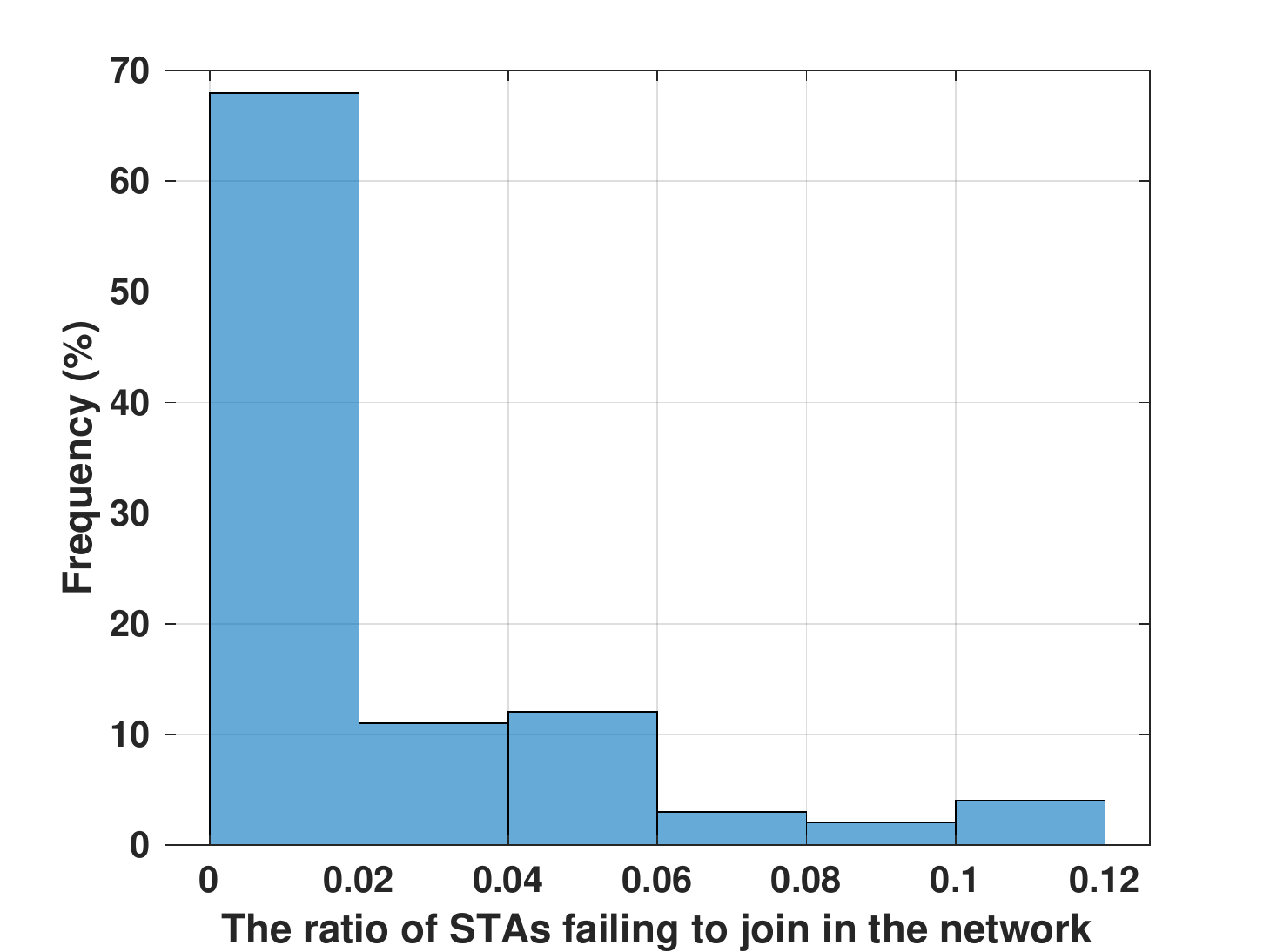}}
	\label{fig:compare}
	\caption{The ratio of STAs failing to join in the network}
\end{figure*}

In real-world scenarios, there always exist some STAs omitted by the CCO due to data frame loss. Therefore, we introduce the robustness mechanism previously mentioned to minimize the number of STAs that fail to connect. To validate the functionality of the robustness mechanism, we conducted repetitive experiments with and without the robust mechanism to construct a 6-layer network, and record the ratios of lost STAs in histograms shown in Fig. 23. The results demonstrated that the ratios of lost STAs were closer to zero with the robust mechanism, while approximately 10\% of STAs were lost without it, highlighting the mechanism's importance in overcoming data frame loss and connecting more STAs.

\section{Conclusion}
In this paper, we concentrate on designing R-PMAC, a preamble based MAC mechanism, which can accelerate the networking of PLC network in IIoT and be robust to collision and data frame loss. In detail, R-PMAC replaces polling with TDMA to save half of time in T-Query and Net-Config, and decreases data frames between CCO and PCO to realize higher time efficiency in multi-layer networking. We propose collision handling mechanism and a retransmission mechanism to eliminate the influence of data frame loss and collision. Our mathematical analysis displays that preamble based MAC machanism is faster than the CSMA used by IEEE1901.1, and R-PMAC is approximately 50\% faster than the P-MAC in \cite{PMAC}. Furthermore, we make a programmable hardware platform and realize the R-PMAC on it. By running the networking of IEEE1901.1, P-MAC and R-PMAC on the hardware platform, we further validate that R-PMAC have better performance than its counterparts. Based on our research, the R-PMAC is a promising candidate for IIoT.

There is still a lot of issues to be studied in the PLC-based IIoT. In \cite{FDPLC}, the idea of using multiple frequency bands was proposed to address the issue of frequent variations in the passband frequency of power lines. Following this, in the future, the PHY layer and MAC layer of PLC can be designed to support multiple central frequencies to make the large-scales more flexibility. In addition, some intermediate layers can be added to the protocol stack of IIoT to facilitate smoother data exchange between lower layers and upper layers. For instance, IPv6 Low Power Wireless Personal Area Network (6LowPAN) \cite{lowPan} is used in wireless IoT to bridge the network layer and link layer. It is expected that the abstract layer techniques like 6LowPAN will play an essential role in the future PLC-based IIoT.

\vspace{12pt}

\end{document}